\documentclass[structabstract]{aa}  

\usepackage[boxruled]{algorithm2e}
\usepackage{epsfig}

\usepackage{txfonts}
\usepackage{natbib}

\newcommand{\nuno}{{\left(\frac{\nu}{\nu_0}\right)}}
\newcommand{\dnuno}{{\left(\frac{\nu-\nu_0}{\nu_0}\right)}}

\newcommand{\X}{\vec{x}}

\newcommand{\B}{\vec{b}}

\newcommand{\V}{\vec{V}}

\newcommand{\A}{{\tens{A}}}
\newcommand{\Ad}{{\tens{A^\dag}}}
\newcommand{\F}{{\tens{F}}}
\newcommand{\Fd}{{\tens{F^\dag}}}
\newcommand{\He}{{\tens{H}}}
\newcommand{\Sa}{{\tens{S}}}
\newcommand{\Sd}{{\tens{S^\dag}}}
\newcommand{\Sna}{\tens{{S_{\nu}}}}
\newcommand{\Snd}{\tens{{S_{\nu}^\dag}}}
\newcommand{\T}{{\tens{T}}}
\newcommand{\W}{{\tens{W}}}

\newcommand{\Pb}{{\vec{P}}}

\newcommand{\Wim}{{{\W^{\rm im}}}}

\newcommand{\Wnt}{{{\W^{\rm {mfs}}_t}}}
\newcommand{\Wntd}{{{{\W^{\rm {mfs}}_t}^\dag}}}

\newcommand{\Wimn}{{{\W^{\rm im}_{\nu}}}}

\newcommand{\wnt}{{w_{\nu}^t}}
\newcommand{\wnq}{{w_{\nu}^q}}
\newcommand{\wntq}{{w_{\nu}^{t+q}}}

\newcommand{\I}{{\vec{I}}}

\newcommand{\Nt}{N_{\rm t}}
\newcommand{\Ns}{N_{\rm s}}
\newcommand{\Nc}{N_{\rm c}}

\begin{document}
%
   \title{A multi-scale multi-frequency deconvolution algorithm
for synthesis imaging in radio interferometry}


   \author{U. Rau
          \inst{1}
          \and
          T.J. Cornwell \inst{2} 
          }

   \institute{ National Radio Astronomy Observatory, Socorro, NM, USA\\
              \email{rurvashi@aoc.nrao.edu}
         \and
             Australia Telescope National Facility, CSIRO, Sydney, AU \\
             \email{tim.cornwell@atnf.csiro.au}
             }

  \date{Received Apr 21, 2011; accepted Jun 04, 2011}

 
  \abstract
   {} 
   { We describe MS-MFS, a multi-scale multi-frequency deconvolution algorithm for
    wide-band synthesis-imaging,  
    and present  
    imaging results that illustrate the capabilities of the algorithm
    and the conditions under which it is feasible and gives accurate results.   
}
   {   
     The MS-MFS algorithm models the wide-band sky-brightness
     distribution as a linear combination of spatial and spectral basis functions, 
     and performs image-reconstruction by combining a linear-least-squares
     approach with iterative $\chi^2$ minimization. 
     This method extends and combines the ideas used
     in the MS-CLEAN and MF-CLEAN algorithms for multi-scale and multi-frequency
     deconvolution respectively, and can be used in conjunction with existing 
    wide-field imaging algorithms.
   We also discuss a simpler hybrid of spectral-line and continuum imaging methods
   and point out situations where it may suffice. 
   }
   { We show via simulations and application to multi-frequency VLA data and
   wideband EVLA data, 
    that it is possible to reconstruct both spatial and spectral structure of 
    compact and extended emission at the continuum sensitivity level and
    at the angular resolution allowed by the highest sampled frequency.
    }
   {}


   \authorrunning{U.Rau and T.J.Cornwell}
   \titlerunning{Multi-Scale Multi-Frequency Synthesis Imaging in Radio Interferometry}
   \maketitle

\section{Introduction}




%
 
Instruments such as the EVLA \citep{EVLA_2009_IEEE}, ASKAP \citep{ASKAP_2009_IEEE} and 
LOFAR \citep{LOFAR_2009_IEEE}
are among a new generation of broad-band radio interferometers,
currently being designed and built to provide
high-dynamic-range imaging capabilities. 
The large instantaneous bandwidths offered by new front-end systems 
increase the raw continuum sensitivity of these instruments 
and 
allow us
to measure the spectral structure 
of the incident radiation across large continuous frequency ranges.
Until recently, the primary goal of wideband imaging 
has been to obtain a continuum image that makes 
use of the increased sensitivity and spatial-frequency 
coverage offered by combining multi-frequency measurements.
So far, 
effects due to the spectral structure of the 
sky-brightness distribution have been considered 
mainly in the context of reducing errors in the continuum image, 
without paying attention to the accuracy of a spectral reconstruction.
But now, the new bandwidths ($100\%$) are large enough to allow the spectral 
structure to also be reconstructed to produce a meaningful astrophysical measurement.   
To do so, 
we need imaging 
algorithms that model and reconstruct both spatial and spectral structure simultaneously,
and that are also 
sensitive to various effects of combining measurements from a large range
 of frequencies
 (namely varying ranges of sampled spatial scales and 
varying array-element response functions).

The simplest method of wide-band image reconstruction is to image 
each frequency channel separately and combine the results at the end. 
However, single-channel imaging is restricted to the narrow-band
$uv$-coverage and sensitivity of the instrument, and source 
spectra can be studied only at the
angular resolution allowed by the lowest frequency in the sampled range.
Also, for complicated extended emission, the single-frequency $uv$-coverage
may not be sufficient to produce a consistent solution across frequency.
While such imaging may suffice for many science goals, it does not
take full advantage of what an instantaneously wide-band instrument provides,
namely the sensitivity and spatial-frequency coverage obtained by 
combining measurements from multiple receiver frequencies during 
image reconstruction.

Multi-Frequency-Synthesis (MFS) \citep{MFCLEAN_CCW}
 is the technique of combining measurements at multiple
discrete receiver frequencies during synthesis imaging.
MFS was initially done  to increase the aperture-plane coverage
of sparse arrays by using narrow-band receivers and switching frequencies during the observations.
Wide bandwidth systems ($\sim10\%$) later presented the problem of bandwidth-smearing, which was
eliminated by splitting the wide band into narrow-band channels and mapping them onto
their correct spatial-frequencies during imaging. It was assumed that
at the receiver sensitivities of the time, 
the measured sky brightness was constant across the observed bandwidth.
The next step was to consider a frequency-dependent sky brightness distribution.
\cite{MFCLEAN_CCW} describe a double-deconvolution algorithm based on
the instrument's responses to a series of spectral basis functions, 
in particular, the first two terms of a Taylor series.
A map of the average spectral index is derived from the coefficient maps.
\cite{MFCLEAN} describe the MF-CLEAN algorithm which uses a formulation
similar to double-deconvolution but calculates Taylor-coefficients via a
least-squares solution.
More recently, \citet{LIKHACHEV} re-derives the least-squares method used
in MF-CLEAN using more than two series coefficients.

%
%

So far, these CLEAN-based multi-frequency deconvolution algorithms used 
point-source (zero-scale) flux components to model the sky emission, 
a choice not well suited for extended emission.
\cite{MSCLEAN} describes the MS-CLEAN algorithm which does
matched-filtering using templates constructed from the instrument 
response to various large scale flux components. 
\cite{ERIC_MSCLEAN} describe a method simular to MS-CLEAN, and 
\cite{Asp_CLEAN} describe the ASP-CLEAN algorithm that explicitly fits 
for the parameters of Gaussian flux components and uses scale size to aid the separation of
signal from noise.
We show in this paper that with the MF-CLEAN approach, 
deconvolution errors that occur with a point-source
model are enhanced in the spectral-index and spectral-curvature images
 because of error propagation effects, and
that the use of a multi-scale technique can minimize this.
%

For high dynamic range imaging across wide fields-of-view,
direction-dependent instrumental effects need to be accounted for.
\cite{AWProjection} describe an algorithm for the correction of time-variable
and wide-field instrumental effects for narrow-band interferometric imaging. 
For wide-field wide-band imaging, these algorithms must be extended to include
the frequency dependence of the instrument. 

In this paper, we describe MS-MFS (multi-scale multi-frequency synthesis) as
an algorithm that combines variants of the MF-CLEAN and MS-CLEAN
approaches to simultaneously reconstruct both spatial and spectral structure of the 
sky-brightness distribution. 
Frequency-dependent primary-beam correction is considered as a 
post-deconvolution correction step\footnote{The integration of 
direction-dependent correction algorithms such as AW-Projection
with MS-MFS will be discussed in a subsequent paper.}.
In section \ref{Sec:IMAGINGRESULTS}, 
we show imaging examples using simulations, multi-frequency VLA data
 and wideband EVLA data,  to illustrate the capabilities and limits of the
MS-MFS algorithm. 


\subsection{Wide-Band Imaging}\label{Sec:WIDEBANDIMAGING}

%
%

We begin with a discussion of how well we can reconstruct both spatial and 
spectral information from an incomplete set of visibilities sampled at
multiple observing frequencies.
An interferometer samples the visibility function of the sky brightness
distribution at a discrete set of spatial-frequencies (called the $uv$-coverage).
The spatial-frequencies sampled at each observing frequency $\nu$ are between  
${u}_{\rm min} = \frac{\nu}{c}{b}_{\rm min} $ and ${u}_{\rm max} = \frac{\nu}{c}{b}_{\rm max}$, 
where ${u}$ is used here as a generic label for the $uv$-distance\footnote
{
The $uv$-distance is defined as $\sqrt{u^2+v^2}$ and is the 
radial distance of the spatial-frequency measured by the baseline from the
 origin of the $uv$-plane, in units of wavelength $\lambda$.
} and 
$b$ represents the length of the baseline vector (in units of meters) projected onto the plane
perpendicular to the direction of the source.
The maximum spatial-frequency measured at each frequency defines the angular 
resolution of the instrument at the frequency ($\theta_{\nu} = 1/u_{\rm max}(\nu)$).
The range of spatial-frequencies between ${u}_{\rm min}$ at $\nu_{\rm max}$ and
${u}_{\rm max}$ at $\nu_{\rm min}$ represents the region where the visibility function
is sampled at all frequencies in the band, and there is sufficient information
to reconstruct both spatial and spectral structure.
The spatial-frequencies outside this region are sampled only by a fraction of the 
band and the accuracy of a broad-band reconstruction
depends on how well the spectral and spatial structure are
constrained by an appropriate choice of flux model. 

Radio interferometric measurements of wideband continuum emission can be
described as one or more of the following situations. 

\begin{enumerate}
\item{Flat Spectrum Sources :}
For a flat-spectrum source, measurements at multiple frequencies sample the 
same  spatial structure, increasing the signal-to-noise of the measurements in 
regions of overlapping spatial-frequencies, and providing better overall 
$uv$-plane filling. The angular resolution of the instrument is given by
$u_{\rm max}$ at $\nu_{\rm max}$.
Standard deconvolution algorithms applied to measurements combined 
via MFS will suffice to reconstruct source structure across the full range of 
spatial scales measured across the band.

%

\item{Unresolved sources with spectral structure : } 
Consider a compact, unresolved source (with spectral structure) that is measured as a point
source at all frequencies. The visibility function of a point source is flat across
the entire spatial-frequency plane. 
Therefore, even if $u_{\rm max}$ changes with frequency, the spectrum of the source 
is adequately sampled by the multi-frequency measurements.
Using a flux model in which each source is a $\delta$-function with a 
specific spectral model (for example, a smooth polynomial), 
it is possible to reconstruct the spectral structure of the source
at the maximum possible angular resolution (given by  $u_{\rm max}$ at $\nu_{\rm max}$).

\item{Resolved sources with spectral structure : }
For resolved sources with spectral structure, the accuracy of the reconstruction
across all spatial scales between $u_{\rm min}$ at $\nu_{\rm min}$ and  $u_{\rm max}$ at $\nu_{\rm max}$
depends on an appropriate choice of flux model, and the constraints that it provides.
For example, a source emitting broad-band synchrotron radiation can be described by
a fixed brightness 
distribution at one frequency with a power-law spectrum associated with each location. 
Images can be made at the maximum angular resolution (given by $u_{\rm max}$ at $\nu_{\rm max}$)
with the 
assumption that different observing frequencies probe the same spatial 
structure but measure different amplitudes (usually a valid assumption). 
This constraint is strong enough to correctly reconstruct even moderately resolved sources
that are completely unresolved at the low end of the band but resolved at the higher end.
On the other hand, a source whose structure itself changes by 100\% in amplitude 
across the band would break the above assumption (band-limited signals).
In this case, a complete reconstruction would be possible only in the region of
overlapping spatial-frequencies (between $u_{\rm min}$ at $\nu_{\rm max}$ and  $u_{\rm max}$ at $\nu_{\rm min}$),
unless the flux model includes constraints that bias the solution towards one
appropriate for such sources.

\item{Spectral structure at large spatial scales : }
The lower end of the spatial-frequency range presents a different problem. 
The size of the central hole in the $uv$-coverage of a typical interferometer 
increases with frequency, and spectra are not measured adequately 
at spatial scales corresponding to spatial-frequencies below 
$u_{\rm min}$ at $\nu_{\rm max}$. 
In the extreme case where most of the visibility function lies within this
$uv$-hole, a flat-spectrum large-scale source (for example) can be indistinguishable 
from a relatively smaller source with a steep spectrum.
Additional constraints in the form of short-spacing spectra  
may be required for an accurate reconstruction.

\item{Frequency dependence of the instrument : }
Array-element responses usually vary with 
frequency, direction and time.  Standard calibration accounts for
the frequency and time dependence for the direction in which the
instrument is pointing. 
Away from the pointing-direction, the frequency-dependent shape 
of the primary-beam is the dominant remaining instrumental effect,
and this results in artificial spectral structure in the images.
To recover both spatial and spectral structure of
the sky brightness across a large field of view, the frequency dependence of
the primary beam must be modeled and removed before or 
during multi-frequency synthesis imaging.  


\end{enumerate}

To summarize, just as standard interferometric image reconstruction uses 
{\it a priori} information about
the spatial structure of the sky to estimate the
visibility function in unmeasured regions of the $uv$-plane,
multi-frequency image reconstruction
algorithms need to use {\it a priori} information about the spectral
structure of the sky brightness. 
By combining a suitable model 
with the known frequency-dependence of the 
spatial-frequency coverage and element response function, 
it is possible to reconstruct the broad-band sky brightness distribution 
from incomplete spectral and spatial-frequency sampling.


%

\section{Multi-scale Multi-frequency deconvolution}\label{Sec:ALGORITHM}

%
The MS-MFS algorithm described here is based on the iterative image-reconstruction
framework
described in \cite{RAU_IEEE_2009} and summarized in Appendix \ref{App:A}. 
Sections \ref{Sec:params} to \ref{Sec:MSMFS} formulate the algorithm
and summarize its implementation in the CASA package.
Differences between the multi-scale and multi-frequency parts of MS-MFS with
the original MF-CLEAN and MS-CLEAN approaches are highlighted in sections
\ref{Sec:relationMF} and \ref{Sec:relationMS}.

\subsection{Parameterization of spatial structure}\label{Sec:params}
An image with multi-scale structure is written as a linear combination
of images at different spatial scales \citep{MSCLEAN}.
\begin{equation}
\vec{I}^{\rm m} = \sum_{s=0}^{\Ns -1}  \vec{I}^{\rm shp}_{s} \star \vec{I}^{\rm sky,\rm{\delta}}_s
\label{Eq:ms_model}
\end{equation}
where $\I^m$ is a multi-scale model image\footnote{
In this paper, superscripts for vectors and matrices indicate type (model,  sky, observed, dirty, residual, etc), and subscripts in italics indicate enumeration indices ($t,q$ for Taylor-term, $s,p$ for spatial scale, $\nu$ for frequency channel.). Non-italic subscripts indicate specific values of the enumerated indices (for example, $\I_{ \rm{\nuup_0} }$,  $\I_{\rm 0}$ or $\I_{\rm\alphaup}$).
}, and 
$\vec{I}^{\rm sky,\rm{\delta}}_{s}$ is a collection of $\delta$-functions that describe the locations
and integrated amplitudes of flux components of scale $s$ in the image.
$\Ns $ is the number of discrete spatial scales used to represent the image  and
$\vec{I}^{\rm shp}_s$ is a tapered truncated parabola of width proportional to $s$.
The symbol $\star$ denotes convolution. 


\subsection{Parameterization of spectral structure}\label{Sec:freqmodel}
The spectrum of each flux component is modeled by a polynomial in frequency
( a Taylor series expansion about $\nu_0$ ).
\begin{eqnarray}
\vec{I}^{\rm m}_{\nu} = \sum_{t=0}^{\Nt -1} \wnt \vec{I}^{\rm sky}_{t} ~~~\mathrm{where}~~~ \wnt&=&\dnuno^t 
\label{Eq:mf_model}
\end{eqnarray}
where $\I^{\rm{\rm sky}}_t$ represents a multi-scale Taylor coefficient image,
and $\Nt $ is the order of the Taylor series expansion.

These Taylor coefficients are interpreted by choosing an astrophysically appropriate
spectral model and performing a Taylor expansion to derive an expression that each coefficient
maps to.
One practical choice is a power law with a varying index, represented by a 
second-order polynomial in $\log(I)~vs~\log\nuno$ space.
\begin{equation}
\I_{\nu}^{\rm sky} = \I_{\nuup_0}^{\rm sky} \nuno^{\I^{\rm sky}_{\alphaup} + \I^{\rm sky}_{\betaup} \log \nuno}
\label{EQN_POWERLAW1}
\end{equation}
Here, $\I^{\rm sky}_{\alphaup}$ represents an average spectral-index, and 
$\I^{\rm sky}_{\betaup}$ represents
spectral-curvature.
The motivation behind this choice of interpretation is the fact that continuum synchrotron emission is 
usually modeled (and observed) as a power law distribution with frequency. Across the wide
frequency ranges that new receivers are now sensitive to, 
spectral breaks, steepening and turnovers need to be factored into
models, and the simplest way to include them and ensure smoothness, is spectral curvature\footnote
{Wideband imaging algorithms described in \cite{MFCLEAN_CCW} and \cite{MFCLEAN} 
use a fixed spectral index across the band,
and handle slight curvature by performing multiple rounds of imaging after removing the 
dominant or average $\alpha$ at each stage. 
They also suggest using higher order polynomials to handle spectral curvature.}.

A Taylor expansion of Eqn.\ref{EQN_POWERLAW1} yields the following expressions for the first
three coefficients from which the spectral index $\I^{\rm sky}_{\alphaup}$ and curvature 
$\I^{\rm sky}_{\betaup}$ images 
can be computed algebraically.
\begin{equation}
\I^m_0 = \I^{\rm sky}_{\nuup_0} ~~;~~ \I^m_1 = \I^{\rm sky}_{\alphaup} \I^{\rm sky}_{\nuup_0} ~~;~~ \I^m_2 = \left(\frac{\I^{\rm sky}_{\alphaup}(\I^{\rm sky}_{\alphaup}-1)}{2} + \I^{\rm sky}_{\betaup}\right) \I^{\rm sky}_{\nuup_0}
\label{EQN_COEFFS}
\end{equation}
Note that with this choice of parameterization, 
we are using a polynomial to model a power-law, and $\Nt $ rapidly increases 
with bandwidth.
A power-series expansion about $\I^{\rm sky}_{\alphaup}$ and $\I^{\rm sky}_{\betaup}$ 
will yield a logarithmic
expansion (i.e. $I$ vs $\log \nu$) which requires fewer coefficients to represent
the same spectrum\footnote{
\citet{MFCLEAN_CCW} state that the logarithmic expansion has better convergence
properties than the linear expansion when $\alpha << 1$.
%
%
An even more compact representation is a polynomial in
$\log I$ vs $\log \nu$, but it becomes numerically unstable to operate on logarithms and 
exponentials of pixel amplitudes, especially in the presence of noise.}.

\subsection{Multi-scale multi-frequency model}
A wideband model of the sky brightness distribution is constructed from 
Eqns \ref{Eq:ms_model} and \ref{Eq:mf_model}.  A wideband flux component
is a spatial basis function ($\I^{\rm shp}_s$, Gaussian or parabola) whose 
integrated amplitude follows a Taylor polynomial in frequency. 
A region of emission in which the spectrum varies with position
will be modeled as a sum of these wide-band flux components.
The image-reconstruction process simultaneously solves for 
spatial and spectral coefficients of these flux components.

The image at each frequency can be modeled as a linear combination of  
Taylor-coefficient images at different spatial scales. 
\begin{equation}
\vec{I}^{\rm m}_{\nu} = \sum_{t=0}^{\Nt } \sum_{s=0}^{\Ns } \wnt \left[ \vec{I}^{\rm shp}_s \star \vec{I}^{\rm sky}_{s\atop t}\right] ~~~~\mathrm{where}~~~\wnt = \dnuno^t 
\label{Eq:msmf_model}
\end{equation}
Here, $\Ns $ is the number of discrete spatial scales used to represent the image and  
$\Nt $ is the order of the series expansion of the spectrum. 
$\vec{I}^{\rm sky}_{s\atop t}$ represents a collection of $\delta$-functions that describe the locations
and integrated amplitudes of flux components of scale $s$ in the image of the $t^{th}$ series 
coefficient. 

\subsection{Measurement equations}\label{Sec:meqn}
The measurement equations\footnote{Appendix \ref{App:A} contains
an explanation of the matrix notation used here, and briefly describes standard
radio-interferometric image-reconstruction within a  
least-squares model-fitting framework (measurement equations, normal equations, and
iterative $\chi^2$ minimization).}
for a sky brightness distribution parameterized by Eqn.\ref{Eq:msmf_model} are

\begin{eqnarray}
\vec{V}^{\rm obs}_{\nu} &=& [\Sna][\F]\vec{I}^{\rm m}_{\nu}
= \sum_{t=0}^{\Nt } \sum_{s=0}^{\Ns } \wnt [\Sna][\T_s][\F] \vec{I}^{\rm sky}_{s\atop t}
\label{Eq:msmfs_meqn}
\end{eqnarray}
$\vec{V}^{\rm obs}_{\nu}$ is a vector of $n\times 1$ visibilities
measured at frequency $\nu$.
$w_{\nu}$ are Taylor-weights (shown in Eqn.\ref{Eq:msmf_model}).
$[\Sna]$ is an $n\times m$ projection operator that represents the 
spatial-frequency sampling function for frequency $\nu$. 
The image-domain convolution of model $\vec{I}^{\rm sky}_{s\atop t}$ with
 $\I^{\rm shp}_s$ is written as a spatial-frequency-domain multiplication.
$ \vec{I}^{\rm shp}_s \star \vec{I}^{\rm sky}_{s\atop t} = [\Fd][\T_s][\F] \vec{I}^{\rm sky}_{s\atop t}$
where $[\T_s]_{m\times m} = diag([\F] \vec{I}^{\rm shp}_s)$ is a spatial-frequency
taper function.  All images are vectors of shape $m\times 1$.

Eqn. \ref{Eq:msmfs_meqn} can be re-written to include all frequencies by
stacking $[\Sna]$ for multiple frequencies. Let $\Nc $ be the number of
frequencies.
\begin{eqnarray}
\vec{V}^{\rm obs} &=& \sum_{t=0}^{\Nt } \sum_{s=0}^{\Ns } [\Wnt][\Sa][\T_s] [\F] \vec{I}^{\rm sky}_{s\atop t}
\label{Eq:msmfs_meqn_2}
\end{eqnarray}
where
$\vec{V}^{\rm obs}$ is a vector of $n\Nc \times 1$ visibilities, 
$[\Sa]$ is an $n\Nc \times m$ sampling matrix representing the multi-frequency
$uv$-coverage of the synthesis array.
$[\Wnt]$ is a diagonal $n\Nc \times n\Nc $ matrix of weights, and 
consists of $\Nc $ diagonal blocks each of size $n\times n$ and containing $\wnt$.

If the summations over $t$ and $s$ are written as a block-matrix dot-product, 
the full measurement matrix has the shape $n\Nc \times m\Ns \Nt $.
When multiplied by the set of $\Ns \Nt $ model sky vectors each
of shape $m\times 1$, it produces $n\Nc $ visibilities.

\noindent For $\Nt =3, \Ns =2$ the measurement equations can be written as follows, in
block matrix form. The subscript $p$ denotes the $p^{th}$ spatial scale
and the subscript $q$ denotes the $q^{th}$ Taylor coefficient of the spectrum
polynomial.

{
\begin{equation}
\begin{array}{l}
\left[\begin{array}{llllll} 
\noalign{\medskip}
\noalign{\medskip}
\noalign{\medskip}
\noalign{\medskip}
   \left[\A_{{0}\atop{0}}\right] & \left[ \A_{{0}\atop{1}} \right] & \left[\A_{{0}\atop{2}}\right] & 
   \left[\A_{{1}\atop{0}}\right] & \left[\A_{{1}\atop{1}}\right] & \left[\A_{{1}\atop{2}}\right] \\  
\noalign{\medskip}
\noalign{\medskip}
\noalign{\medskip}
\noalign{\medskip}
   \end{array} \right] \\
\noalign{\medskip}
\noalign{\medskip}
\noalign{\medskip}
\noalign{\medskip}
\noalign{\medskip}
{~\mathrm{where}~~\left[ \A_{{p}\atop{q}}\right] = [\W^{\rm mfs}_q][\Sa][\T_p][\F]} \\
\noalign{\medskip}
{~\mathrm{for}~p\in\{0,\Ns -1\}~\mathrm{and}~q\in\{0,\Nt -1\}    }  
\end{array}
\left[\begin{array}{l} 
\noalign{\medskip}
                       \vec{I}^{\rm sky}_{{0}\atop{0}} \\ 
\noalign{\medskip}
                       \vec{I}^{\rm sky}_{{0}\atop{1}} \\ 
\noalign{\medskip}
                       \vec{I}^{\rm sky}_{{0}\atop{2}} \\ 
\noalign{\medskip}
                       \vec{I}^{\rm sky}_{{1}\atop{0}} \\ 
\noalign{\medskip}
                       \vec{I}^{\rm sky}_{{1}\atop{1}} \\ 
\noalign{\medskip}
                       \vec{I}^{\rm sky}_{{1}\atop{2}}\\
\noalign{\medskip}
		       \end{array}\right] =\vec{V}^{\rm obs}
\label{meqn_msmfs_math}
\end{equation}
}

\subsection{Normal equations}\label{Sec:neqn}

The normal equations 
for the system described in Eqn. \ref{Eq:msmfs_meqn}
can be written in block matrix form, with each block-row 
(for scale size $s$, and Taylor term $t$) given by
\begin{eqnarray}
\label{Eq:msmfs_neqn_1}
\sum_{p=0}^{\Ns -1}\sum_{q=0}^{\Nt -1} \left[\He_{{s,p}\atop{t,q}}\right] \vec{I}^{\rm sky}_{{p}\atop{q}} &=& \vec{I}^{\rm dirty}_{{s}\atop{t}}~~~  \forall~ s \in [0,\Ns -1], t\in[0,\Nt -1]
\end{eqnarray}
Here, each $\left[\He_{{s,p}\atop{t,q}} \right]$ is an $m\times m$ block of the 
Hessian matrix, and $\vec{I}^{\rm dirty}_{{s}\atop{t}}$ is one of $\Ns  \Nt $ dirty images.
\begin{eqnarray}
\left[\He_{{s,p}\atop{t,q}} \right] &=& \left[\A_{{s}\atop{t}}^{\dag}\right][\Wim] \left[\A_{{p}\atop{q}}\right] \\
 &=&    [\Fd \T_s  \Sd \Wntd ]  [\Wim]  [\W^{\rm mfs}_q \Sa \T_p \F] \\
 &=&    [\Fd \T_s \F] [\Fd  \Sd \Wntd  \Wim  \W^{\rm mfs}_q \Sa  \F] [\Fd \T_p \F]\\
 &=& [\Fd \T_s \F] \left\{  \sum_{\nu} \wntq [\Fd\Snd\Wimn\Sna\F] \right\} [\Fd \T_p \F] \\
 &=& [\Fd \T_s \F] \left\{  \sum_{\nu} \wntq [\He_{\nu}\} \right\} [\Fd \T_p \F] 
\end{eqnarray}
$[\Wim]$ is a diagonal matrix of data-weights (and imaging-weights) and  
$[\W^{\rm mfs}_t]$ is a diagonal matrix containing Taylor-weights $\wnt$.
$[\He_{\nu}] =  [\Fd\Snd\Wimn\Sna\F]$ is the Hessian matrix formed using only one
 frequency channel, and
is a convolution operator containing a shifted version of the single-frequency 
point-spread-function $\I^{\rm psf}_{\nu} = diag[\Fd \Sd \Wimn \Sa]$ in each row 
(see appendix \ref{App:normal} for details).
$[\Fd \T_s\F]$ and $[\Fd\T_p\F]$ are also convolution operators 
with $\I^{\rm shp}_s$ and $\I^{\rm shp}_p$ as their kernels. 
The process of convolution is associative and commutative, and 
therefore, $\left[\He_{{s,p}\atop{t,q}}\right]$  is also a convolution operator 
whose kernel is given by 
\begin{equation}
\label{Eq:msmfs_neqn_2.5}
\vec{I}^{\rm psf}_{{s,p}\atop{t,q}} = \I^{\rm shp}_s  \star \left\{ \sum_{\nu} \wntq \vec{I}^{\rm psf}_{\nu} \right\} \star \I^{\rm shp}_p 
\end{equation}


\noindent The dirty images on the RHS of Eqn.\ref{Eq:msmfs_neqn_1} can be written as follows.
\begin{eqnarray}
\label{Eq:msmfs_neqn_3}
\vec{I}^{\rm dirty}_{{s}\atop{t}}  &=&  \left[\A_{{s}\atop{t}}^{\dag}\right][\Wim]  \vec{V}^{\rm obs}\\
&=& [\Fd \T_s \Sd\Wntd\Wim] \vec{V}^{\rm obs} \\
&=& [\Fd \T_s \F][\Fd\Sd\Wntd\Wim] \vec{V}^{\rm obs} \\
&=& [\Fd \T_s \F] \left\{  \sum_{\nu} \wnt [\Fd\Snd\Wimn] \vec{V}_{\nu}^{\rm obs}  \right\} \\
&=& \I^{\rm shp}_s \star \left\{\sum_{\nu} \wnt \vec{I}^{\rm dirty}_{\nu}  \right\} 
\label{Eq:msmfs_neqn_3a}
\end{eqnarray}
where ${\I}_{\nu}^{\rm dirty} = [\Fd \Snd \Wimn] \vec{V}^{\rm obs}_{\nu} $ is the dirty
image formed by direct Fourier inversion of weighted visibilities from one frequency channel.

When all scales and Taylor terms are combined, 
the full Hessian matrix contains
$\Nt  \Ns  \times \Nt  \Ns $ blocks each of size $m\times m$, 
and $\Nt $ Taylor coefficient images each of size $m\times 1$, 
for all $\Ns $ spatial scales.

The normal equations in block matrix form for the example 
in Eqn.\ref{meqn_msmfs_math} for $\Nt =3, \Ns =2$ is shown
in Eqn.\ref{Eq:msmfs_neqn_matrix}.
The Hessian matrix consists of $\Ns \times \Ns =2\times 2$ blocks 
(the four quandrants of the matrix), each for one pair of spatial scale $s,p$ (the upper indices).
Within each quadrant, the $\Nt \times \Nt  = 3\times 3$
matrices correspond to various pairs of $t,q$ (Taylor coefficient indices; the lower indices).
This layout shows how the multi-scale and multi-frequency aspects of
this imaging problem are combined and illustrates the 
dependencies between the spatial and spectral basis functions.

\begin{equation}\small
\left[\begin{array}{llllll} 
\noalign{\medskip}
   \left[\He_{{ 0, 0}\atop{ 0, 0}}\right] & \left[\He_{{ 0, 0}\atop{ 0, 1}}\right] & \left[\He_{{ 0, 0}\atop{ 0, 2}}\right] & \left[\He_{{ 0, 1}\atop{ 0, 0}}\right] & \left[\He_{{ 0, 1}\atop{ 0, 1}}\right] & \left[\He_{{ 0, 1}\atop{ 0, 2}}\right] \\  
\noalign{\medskip}
   \left[\He_{{ 0, 0}\atop{ 1, 0}} \right] & \left[\He_{{ 0, 0}\atop{ 1, 1}}\right] & \left[\He_{{ 0, 0}\atop{ 1, 2}}\right] & \left[\He_{{ 0, 1}\atop{ 1, 0}}\right] & \left[\He_{{ 0, 1}\atop{ 1, 1}}\right] & \left[\He_{{ 0, 1}\atop{ 1, 2}}\right] \\  
\noalign{\medskip}
   \left[\He_{{ 0, 0}\atop{ 2, 0}} \right] & \left[\He_{{ 0, 0}\atop{ 2, 1}}\right] & \left[\He_{{ 0, 0}\atop{ 2, 2}}\right] & \left[\He_{{ 0, 1}\atop{ 2, 0}}\right] & \left[\He_{{ 0, 1}\atop{ 2, 1}}\right] & \left[\He_{{ 0, 1}\atop{ 2, 2}}\right] \\  
\noalign{\medskip}
\noalign{\medskip}
   \left[\He_{{ 1, 0}\atop{ 0, 0}} \right] & \left[\He_{{ 1, 0}\atop{ 0, 1}}\right] & \left[\He_{{ 1, 0}\atop{ 0, 2}}\right] & \left[\He_{{ 1, 1}\atop{ 0, 0}}\right] & \left[\He_{{ 1, 1}\atop{ 0, 1}}\right] & \left[\He_{{ 1, 1}\atop{ 0, 2}}\right] \\  
\noalign{\medskip}
   \left[\He_{{ 1, 0}\atop{ 1, 0}} \right] & \left[\He_{{ 1, 0}\atop{ 1, 1}}\right] & \left[\He_{{ 1, 0}\atop{ 1, 2}}\right] & \left[\He_{{ 1, 1}\atop{ 1, 0}}\right] & \left[\He_{{ 1, 1}\atop{ 1, 1}}\right] & \left[\He_{{ 1, 1}\atop{ 1, 2}}\right] \\  
\noalign{\medskip}
   \left[\He_{{ 1, 0}\atop{ 2, 0}} \right] & \left[\He_{{ 1, 0}\atop{ 2, 1}}\right] & \left[\He_{{ 1, 0}\atop{ 2, 2}}\right] & \left[\He_{{ 1, 1}\atop{ 2, 0}}\right] & \left[\He_{{ 1, 1}\atop{ 2, 1}}\right] & \left[\He_{{ 1, 1}\atop{ 2, 2}}\right] \\  
\noalign{\medskip}
   \end{array} \right]
\left[\begin{array}{l} 
\noalign{\medskip}
		       \vec{I}^{\rm sky}_{{ 0}\atop{ 0}} \\ 
\noalign{\medskip}
                       \vec{I}^{\rm sky}_{{ 0}\atop{ 1}}\\ 
\noalign{\medskip}
		       \vec{I}^{\rm sky}_{{ 0}\atop{ 2}}\\ 
\noalign{\medskip}
\noalign{\medskip}
		       \vec{I}^{\rm sky}_{{ 1}\atop{ 0}} \\ 
\noalign{\medskip}
		       \vec{I}^{\rm sky}_{{ 1}\atop{ 1}}\\ 
\noalign{\medskip}
		       \vec{I}^{\rm sky}_{{ 1}\atop{ 2}}\\
\noalign{\medskip}
			       \end{array}\right] =
\left[\begin{array}{l} 
\noalign{\medskip}
                       \vec{I}^{\rm dirty}_{{ 0}\atop{ 0}}  \\ 
\noalign{\medskip}
                       \vec{I}^{\rm dirty}_{{ 0}\atop{ 1}} \\ 
\noalign{\medskip}
		       \vec{I}^{\rm dirty}_{{ 0}\atop{ 2}}\\ 
\noalign{\medskip}
\noalign{\medskip}
		       \vec{I}^{\rm dirty}_{{ 1}\atop{ 0}} \\ 
\noalign{\medskip}
		       \vec{I}^{\rm dirty}_{{ 1}\atop{ 1}} \\ 
\noalign{\medskip}
		       \vec{I}^{\rm dirty}_{{ 1}\atop{ 2}} \\
\noalign{\medskip}
			       \end{array}\right] 
\label{Eq:msmfs_neqn_matrix}
\end{equation}

This is the system of equations to be solved.
The spatial-frequency sampling  of a real interferometer is always incomplete 
($[\Sa]$ is rank-deficient).
Therefore, each Hessian block, and the entire Hessian matrix is singular, and an exact inverse
does not exist\footnote
{Even if $[\He]$ were invertible, it is impractical to evaluate and invert the full Hessian
(each row of each Hessian block represents an image).
}. An accurate reconstruction can be obtained only via successive approximation
(iterative numerical optimization).

\subsection{Principal Solution}\label{Sec:prinsol}
The principal solution\footnote{The principal solution 
(as defined in \citet{BRACEWELL_ROBERTS} and used in 
\citet{DECONV_LECTURE}) is a term specific to radio interferometry
and represents the dirty image normalized by the sum of weights.  
It is the image formed purely from the measured data, with no contribution from the
invisible distribution of images (unmeasured spatial-frequencies).
It is also an approximate solution of the normal equations $[\He] I^{\rm sky}=I^{\rm dirty}$
(see Appendix \ref{App:normal}), 
calculated using a diagonal approximation of the Hessian.
Each element on the diagonal  is the peak of the PSF, which is also the
sum of weights.
For isolated sources, the values measured at the peaks of the
principal solution (dirty) images are the true sky values. 
The CLEAN minor-cycle algorithm uses this fact to estimate source
fluxes from the peaks of the normalized dirty image.
} of the normal equations is an approximate solution that can be
computed via diagonal approximations of all Hessian blocks.
This solution is then used to pick flux components
 within the minor-cycle of iterative deconvolution.

Each Hessian block is a convolution operator with a shifted version of
a point-spread-function $\I^{\rm psf}_{{s,p}\atop{t,q}}$ in
each row (their centers are aligned on the diagonal). 
A diagonal approximation represents the assumption that the PSFs are $\delta$-functions, or that
the amplitudes in the dirty image at the location of a source reflect the true flux of the source.
Also, with the assumption of spatially invariant PSFs, all elements on the diagonal within
each Hessian block are the same. Therefore, we can reduce
 each $\left[\He_{{s,p}\atop{t,q}}\right]$
to one number. The full Hessian reduces to an 
%
%
$\Nt \Ns  \times \Nt \Ns $ element matrix $[\He^{\rm peak}]$, 
The principal solution 
can be obtained by inverting $[\He^{\rm peak}]$ once, and applying it to the dirty 
image vectors, one pixel at a time.
Such a solution will be correct only at the locations of the centers
of isolated flux-components and must be augmented with an iterative optimization 
approach to ensure accuracy.  In the case of perfect sampling (where the Hessian blocks
are truly diagonal and PSFs are $\delta$-functions), the principal solution will directly
give correct images of Taylor-series coefficients.

%

\subsubsection{Properties of $[{\tens{H}}^{\rm peak}]$}
Some properties of $[\He^{\rm peak}]$ are worth noting, to understand the numerical stability
of this approach and its dependence on the choice of spectral and spatial basis functions
(sky model), and spectral and spatial-frequency sampling functions (data and instrument). 

\begin{enumerate}
\item There are $\Ns \Nt $ elements on the diagonal of $[\He^{\rm peak}]$.
Each is a measure of the instrument's sensitivity 
to a flux component of unit total flux whose shape and spectrum are described by one
pair of spatial and spectral basis functions ($\I^{\rm shp}_s$ and $\dnuno^t$).
Each diagonal element is given by
\begin{eqnarray}
H^{\rm peak}_{{s,s}\atop{t,t}} &=& mid\left\{\vec{I}^{\rm psf}_{{s,s}\atop{t,t}}\right\} = tr\left[\sum_{\nu} w_{\nu}^{t+t} [\T_s\Snd\Wimn\Sna \T_s] \right] \\
	& &\forall~ s ~\in ~\{0 ... \Ns -1\}~~,~~t ~\in ~\{0 ... \Nt -1\}\nonumber
\label{Eq:hpeak_msmfs_1}
\end{eqnarray}
Note that the instrument's sampling function and data-weights are included in this expression.

\item The off-diagonal elements measure the orthogonality\footnote
{\label{FN:ortho}The following definition of orthogonality is used here. Two vectors are orthogonal
if their inner product is zero. The orthogonality of a pair of scale functions 
is measured by the integral of the product of their $uv$-taper functions.
To account for $uv$-coverage, this integral is weighted by the sampling function.
}
 between the various
basis functions, for the given $uv$-coverage and weighting scheme.
They measure the amount of overlap between basis
functions in the measurement domain. 
Smaller values indicate a more orthogonal set of basis functions
that the instrument is better able to distinguish between. 

\item The condition number of this matrix (or of blocks within this matrix)
will indicate if the chosen set of basis functions and spatial-frequency coverage 
provide enough constraints to provide a stable solution, and 
can be used as a metric to choose a suitable basis set.
For a simple example, if a 3-term solution is attempted with data from only two distinct
frequencies, $[\He^{\rm peak}]$ will be singular.
Or, for some choice of multi-frequency $uv$-coverage, the visibilities
measured by the instrument for two different spatial scales may become hard to distinguish. 
Then, the cross-term element of $[\He^{\rm peak}]$ corresponding to
this combination could have a higher value, indicating that the two parameters are highly coupled,
and there is insufficient information in the data and sampling pattern to distinguish 
between the scales. A similar situation can arise to create ambiguity between 
spatial and spectral structure (an extreme example is multi-frequency measurements from only
one baseline).
\item In general, $[\He^{\rm peak}]$ will be a positive-definite 
symmetric matrix whose inverse can be easily computed {\it via} a 
Cholesky decomposition\footnote
{A Cholesky decomposition factors a symmetric positive-definite matrix
into a lower triangular matrix and its conjugate transpose. 
It is used in the solution of system of equations $[A]\vec{x}=\vec{b}$ where
$[A]$ is symmetric positive-definite. 
The normal equations of a linear least-squares problem in which the signal is modeled
as a linear combination of basis functions, are usually in this form  
\citep{NR}.
}.  
The value of $\Ns $ is usually $< 10 $, making the inversion 
of $[\He^{\rm peak}]$ tractable as a one-time operation.

\item 
Some further approximations can be made about the structure
of $[\He^{\rm peak}]$ to simplify its inversion, and it is 
important to understand the numerical implications of these trade-offs.
One is a block-diagonal approximation of $[\He^{\rm peak}]$ 
({\it i.e.} using only those blocks of the Hessian in
Eqn.~\ref{Eq:msmfs_neqn_matrix} for which $s=p$; top-left
and bottom-right quadrants).
This approximation treats each spatial scale separately and
assumes that the scale basis functions are orthogonal.
For each scale, cross-terms between Taylor functions are preserved, and
a multi-frequency principal solution can be obtained separately for each spatial scale.
Note that a set of tapered truncated paraboloids is never orthogonal, but this 
separate-scale approximation
works because of the iterative $\chi^2$-minimization process.
This approximation makes the Hessian inversion easier, but to preserve
accuracy, the update step of the iterative deconvolution still needs to evaluate
the full LHS of the normal equations while subtracting out a flux component.

\end{enumerate}

\subsection{MS-MFS algorithm}\label{Sec:MSMFS}

This section describes an iterative process that 
solves the normal equations (Eqn.\ref{Eq:msmfs_neqn_1}) and produces a
set of $\Nt $ Taylor-series coefficient images at $\Ns $ different spatial scales.
Appendix \ref{App:algolisting} 
lists the algorithm-steps in pseudo-code format, reflecting the implementation
of MS-MFS in the CASA\footnote{\href{http://casa.nrao.edu}
{Common Astronomy Software Applications} is being developed at the
\href{http://www.nrao.edu}{National Radio Astronomy Observatory}} 
 software package.

\noindent\paragraph{\bf Pre-compute Hessian : }
Each block of the Hessian is a convolution operator, consisting of a shifted version 
of the same convolution kernel in each row. Therefore, it suffices to compute and
store one kernel per Hessian block. 
Convolution kernels for all distinct blocks in the $\Ns  \Nt  \times \Ns  \Nt $ Hessian 
are evaluated {\it via} Eqn.~\ref{Eq:msmfs_neqn_2.5}. 
All kernels are normalized by the sum-of-weights such that the peak of $\vec{I}^{\rm psf}_{{0,0}\atop{0,0}}$
is unity, and the relative weights between Hessian blocks is preserved. 
A block-diagonal approximation of $[\He^{\rm peak}]$ is done, and a 
set of $\Ns $ matrices each of shape $\Nt \times \Nt $ and denoted as $[\He^{\rm peak}_s]$
are constructed.
Their inverses are computed and stored in $[{\He^{\rm peak}_s}^{-1}]$.


\noindent\paragraph{\bf Initialization :} 
All $\Ns \Nt $ model images are initialized to zero (or an {\it a priori} model to start from).

\noindent\paragraph{\bf Major and minor cycles : }
Iterative image reconstruction in radio interferometry is usually split into major and minor
 cycles (see Appendix \ref{App:iterations}).
The major cycles compute the RHS of the normal equations, and the 
minor cycle inverts
the Hessian and gets an estimate of the model. This model is used in the next major cycle to
compute new RHS vectors from the residuals, and the process repeats itself
until convergence is achieved.
%
Steps \ref{stepRHS_msmfs} and \ref{stepPredict_msmfs} (of the list of steps given below) 
form one major cycle, 
and repetitions of Steps \ref{stepFind_msmfs} through \ref{stepUpdate_2_msmfs} form the minor cycle.
\begin{enumerate}
\item\label{stepRHS_msmfs} {\bf Compute residual images :} The RHS vectors 
(residual or dirty images) $\vec{I}^{\rm dirty}_{{s}\atop{t}} ~\forall~t\in\{0,\Nt $-$1\}$
of the normal equations are computed {\it via} Eqn.\ref{Eq:msmfs_neqn_3a} by first
computing the multi-frequency dirty images and then convolving them by the
scale basis functions.

\item\label{stepFind_msmfs} {\bf Find a Flux Component :}
The principal solution is computed for all pixels, one scale at a time.
\begin{equation}
\I^{\rm pix,psol}_s = [{\He^{\rm peak}_s}^{-1}] \I^{\rm pix,dirty}_s ~~~~~~~\mathrm{for~each~pixel,~and~scale} ~s
\label{Eq:msmfs_psol}
\end{equation}
Here, 
$\I^{\rm pix,psol}_s$ is a list of $\Nt $ Taylor-coefficients for the pixel-location $pix$ and scale $s$.
$[\He^{\rm peak}_s]$ is the $s^{th}$ block (of size $\Nt \times \Nt $)
in the list of diagonal-blocks of $[\He^{\rm peak}]$, 
and $\I^{\rm pix,dirty}_s$ is the $\Nt \times 1$ vector constructed 
from $\vec{I}^{\rm dirty}_{{s}\atop{t}} ~\forall~t\in\{0,\Nt $-$1\}$ for one pixel $pix$. 
This step is performed on all pixels, separately for all scales $s$, resulting in  $\Ns $ sets of 
$\Nt $ Taylor-coefficient images.
The most appropriate set of Taylor coefficients must now be chosen.
The search is performed across pixels and scales. Many heuristics can be used here.
For example, in iteration $i$,
choose the $\Nt $ element solution-set with the dominant $q=0$ component 
across all scales and pixel locations. 
Or, pick the set of components that makes the
largest impact on the value of $\chi^2$. 
Or, choose the location from the peak in the $t=0$ residual image, and compute
the principal solution only for that pixel.

The result of this step is a set of $\Nt $ model images, each containing one 
$\delta$-function that marks the location of the center of a flux component
of shape $\I^{\rm shp}_{p,(i)}$  
($p$ represents the scale of the chosen component, out of all possible values of $s$).
The amplitudes of these $\Nt $ $\delta$-functions
are the Taylor coefficients that model the spectrum of the integrated flux of 
this component. Let these $\Nt $ model images from iteration $i$ be denoted as 
$\left\{\vec{I}^{\rm m}_{{{p}\atop{q}},(i)}\right\};q\in[0,\Nt ]$.

\item\label{stepUpdate_msmfs} {\bf Update model images :}
A set of $\Nt $ multi-scale model images are accumulated.
\begin{equation}
\I^{\rm m}_{q} = \I^{\rm m}_{q} + g \left( \I^{\rm m}_{{{p}\atop{q}},(i)} \star \I^{\rm shp}_{p} \right) ~~~~~~~~ \forall q\in[0,\Nt ]
\label{Eq:msmfs_updatemodel}
\end{equation}
where $g$ is a loop-gain that takes on values between 0 and 1 and controls the 
step size for each iteration in the $\chi^2$-minimization process.

\item\label{stepUpdate_2_msmfs} {\bf Update RHS :}
The RHS residual images\footnote{
In the first iteration, the RHS vectors are called the dirty-images. In all subsequent
steps, the RHS vectors are formed after subtracting model-visibilities from the data
and are called residual-images.
} are updated by evaluating and subtracting out the entire
LHS of the normal equations. 
%
%
However, since the chosen flux component corresponds
to just one scale, this becomes a summation over only Taylor terms (and not scale; there is
only one scale $p$ in the current model).
\begin{equation}
\I^{\rm res}_{{s}\atop{t}} = \I^{\rm res}_{{s}\atop{t}} - g \sum_{q_i=0}^{\Nt -1} \left[  \I^{\rm psf}_{{s,p}\atop{t,q}}  \star \I^{\rm m}_{{{p}\atop{q}},(i)} \right] 
\label{Eq:msmfs_updaterhs}
\end{equation}
{\bf Repeat from Step \ref{stepFind_msmfs}} until the minor-cycle flux limit is reached.

\item\label{stepPredict_msmfs} {\bf Predict }:
Model visibilities are computed from the set of Taylor-coefficient images via Eqn.\ref{Eq:msmfs_meqn}.
Residual visibilities are computed as 
$\vec{V}^{\rm res}_{\nu}= \vec{V}^{\rm obs}_{\nu} - \vec{V}^{\rm m}_{\nu}$.

{\bf Repeat from Step \ref{stepRHS_msmfs}} until a global convergence criterion is satisfied.
(After the first iteration, $\vec{V}^{\rm res}_{\nu}$ is used as the new $\vec{V}^{\rm obs}_{\nu}$ in Step \ref{stepRHS_msmfs}).
\end{enumerate}

\noindent\paragraph{\bf Restoration :}
After convergence, the model Taylor-coefficient images can be interpreted in
different ways. 
\begin{enumerate}
\item The most obvious data products are the Taylor-coefficient images themselves, 
which are directly smoothed by the restoring beam. Residual images are added
back in after computing the principal solution from the residuals obtained in 
the last instance of Step \ref{stepRHS_msmfs}, to
ensure that any undeconvolved flux has the correct flux values\footnote
{As pointed out in section \ref{Sec:prinsol}, this will be accurate only for isolated
point sources that were left out of the minor cycle.
}.
\item  For the study of broad-band radio emission, the spectral coefficients can be
interpreted in terms of a power law in frequency with varying index
(as described in Section \ref{Sec:freqmodel}).
The data products are images of the reference-frequency flux $\vec{I}^{\rm m}_{\nuup_0}$,
the spectral-index $\vec{I}^{\rm m}_{\alpha}$ and the spectral curvature $\vec{I}^{\rm m}_{\betaup}$.
\begin{eqnarray}
\label{Eq:calcab_1}
\vec{I}^{\rm m}_{\nuup_0} &=& \vec{I}^{\rm m}_0 \\
\label{Eq:calcab_2}
\vec{I}^{\rm m}_{\alphaup} &=& {\vec{I}^{\rm m}_1}/{\vec{I}^{\rm m}_0}  \\
\label{Eq:calcab_3}
\vec{I}^{\rm m}_{\betaup} &=& \left[{\vec{I}^{\rm m}_2}/{\vec{I}^{\rm m}_0}\right] - \left[{{\vec{I}^{\rm m}_{\alphaup}(\vec{I}^{\rm m}_{\alphaup}-1)}}{/2}\right]
\end{eqnarray}
Spectral index and curvature images are calculated only in regions where
the values in $\vec{I}^{\rm m}_0$ are above a chosen threshold.
\item An image cube can be constructed by evaluating the spectral polynomial
{\it via} Eqn.~\ref{Eq:mf_model} for each frequency. 
This data product is
useful for sources whose emission is not well modeled by a power law,
but is a smooth polynomial in frequency. Band-limited signals that taper off
smoothly in frequency are one example.
\item 
When multiple sources along a given line-of-sight have different spectra, the 
Taylor-coefficients will represent the combined spectrum. 
To compute spectral index and curvature maps for foreground sources,
a polynomial background-subtraction must be done on the Taylor-coefficient images
before Eqns\ref{Eq:calcab_1} to \ref{Eq:calcab_3} are evaluated.
\end{enumerate}

\noindent \paragraph{\bf Primary-beam correction : } 
For wide-field imaging, the spatial and spectral structure of the primary beam
(array-element response function)
contributes to the measured signal. If this instrumental effect is not accounted for, 
the output Taylor-coefficient images approximately 
represent the product of the sky and the primary-beam. 
\begin{equation}
\I_{\nu}^{\rm sky} = \Pb_{\nu}  \I_{\nu}^{\rm true} =  \Pb_{\nuup_0} \I_{\nuup_0}^{\rm true} \nuno^{[\I^{\rm true}_{\alphaup}+\vec{P}_{\alphaup}] + [\I^{\rm true}_{\betaup} + \vec{P}_{\betaup} ] \log \nuno}
\label{EQN_POWERLAW2}
\end{equation}
$\vec{P}_{\nuup_0}$ is the primary beam at the reference frequency, and
$\vec{P}_{\alphaup}$ and $\vec{P}_{\betaup}$ are spectral index and curvature due to the frequency
dependence of the primary beam.  

A correction for the average primary-beam and its frequency dependence can be
done as a post-deconvolution step. The primary-beams are first evaluated or measured 
as a function of frequency, and the frequency-dependence per pixel modeled by a
power-law or a polynomial (perferably the same spectral polynomial used for the image
reconstruction). Primary-beam correction can then be done as follows.
\begin{eqnarray}
\vec{I}^{\rm new}_{\nuup_0}&=&\vec{I}^{\rm m}_{\nuup_0}/\vec{P}_{\nuup_0}\\
 \vec{I}^{\rm new}_{\alphaup}&=&\vec{I}^{\rm m}_{\alphaup}-\vec{P}_{\alphaup}\\
 \vec{I}^{\rm new}_{\betaup}&=&\vec{I}^{\rm m}_{\betaup}-\vec{P}_{\betaup}
\end{eqnarray}
Note that if a polynomial is fit for the frequency-dependence of the primary beam, 
and $\vec{P}_{\nuup_0}, \vec{P}_{\alphaup}, \vec{P}_{\betaup}$ computed from it, the above 
operation is numerically identical to doing a polynomial division in terms of two sets of
coefficients (for $\Nt <=3$). A brute-force polynomial-division using more series coefficients
will yield a more accurate solution.   
Note however, that such a correction will not be
accurate if there are time-dependent variations in the primary-beam, and will require
integration with the AW-Projection algorithm discussed in \cite{AWProjection}.

%
%
%
%

\section{Relation to MF-CLEAN and MS-CLEAN}
The MS-MFS algorithm is a combination of the general ideas used in MS-CLEAN and MF-CLEAN,
but there are some subtle differences. The next two sections briefly discuss these differences
and their numerical implications.

\subsection{Relation to MF-CLEAN}\label{Sec:relationMF}
The MF-CLEAN algorithm \citep{MFCLEAN} models the sky as a collection of point sources
with a Taylor polynomial spectrum. A point-source version of MS-MFS can be derived by setting
$\Ns =1$ and using $\I^{\rm shp}_0 = \delta$-function in the derivations in 
sections \ref{Sec:meqn} and \ref{Sec:neqn}.
%
%
\noindent The normal equations can be written in block matrix form (for example, for $\Nt =3$).
\begin{equation}
\left[\begin{array}{lll} 
   [\He_{0,0}] & [\He_{0,1}] & [\He_{0,2}]\\  
\noalign{\medskip}
   [\He_{1,0}] & [\He_{1,1}] & [\He_{1,2}] \\  
\noalign{\medskip}
   [\He_{2,0}] & [\He_{2,1}] & [\He_{2,2}] \\  
   \end{array} \right]
\left[\begin{array}{l} \vec{I}^{\rm sky}_{0} \\ 
\noalign{\medskip}
                       \vec{I}^{\rm sky}_{1} \\ 
\noalign{\medskip}
		       \vec{I}^{\rm sky}_{2}\end{array}\right] =
\left[\begin{array}{l} \vec{I}^{\rm dirty}_{0}\\
\noalign{\medskip}
		       \vec{I}^{\rm dirty}_{1} \\
\noalign{\medskip}
		       \vec{I}^{\rm dirty}_{2}\end{array}\right] 
\label{Eq:mfs_neqn3}
\end{equation}
Each block $[\He_{t,q}]$ is a convolution operator with $\vec{I}^{\rm psf}_{t,q}$ as its kernel.
\begin{eqnarray}
\label{Eq:mf_neqn_2b}
\vec{I}^{\rm psf}_{t,q}  &=& \sum_{\nu} \wntq \vec{I}^{\rm psf}_{\nu}\\
\label{Eq:mf_neqn_3a}
\vec{I}^{\rm dirty}_{t}  &=& \sum_{\nu} \wnt \vec{I}^{\rm dirty}_{\nu}
\end{eqnarray}
On the other hand, the MF-CLEAN algorithm described in \cite{MFCLEAN} follows a
matched-filtering approach using functions called spectral-PSFs, 
which are equivalent to the convolution kernels from the first row
of Hessian blocks ($q=0$) in Eqn.\ref{Eq:mf_neqn_2b}.
In MF-CLEAN, the Hessian elements and RHS vectors are calculated by convolving spectral-PSFs with 
themselves and the residual images.
\begin{eqnarray}
\label{Eq:mfclean_2}
\vec{I}^{\rm psf}_{t,q}  &=& \left\{ \sum_{\nu}\wnt \vec{I}^{\rm psf}_{\nu}  \right\} \star \left\{ \sum_{\nu}\wnq \vec{I}^{\rm psf}_{\nu}  \right\}\\
\label{Eq:mfclean_3}
\vec{I}^{\rm dirty}_{t}  &=& \left\{ \sum_{\nu}\wnt \vec{I}^{\rm psf}_{\nu}  \right\} \star \left\{\sum_{\nu} \vec{I}^{\rm dirty}_{\nu}  \right\}
\end{eqnarray}
Formally, this matched filtering approach is exactly equal to the
calculations shown in 
Eqns \ref{Eq:mf_neqn_2b} and \ref{Eq:mf_neqn_3a} {\it only} under the conditions that there
is no overlap on the spatial-frequency plane between measurements from different
observing frequencies, and all measurements are weighted equally across the spatial-frequency
plane (uniform weighting).
%
In practice, MF-CLEAN incurs errors for arrays with dense  
spatial-frequency coverage where tracks from different baselines and frequency-channels
intersect.
When applied to simulated EVLA data, numerical
instabilities limited the fidelity of the final image, especially with
extended emission, and this instability was eliminated by changing the computations
from Eqns \ref{Eq:mfclean_2} and \ref{Eq:mfclean_3}
to Eqns \ref{Eq:mf_neqn_2b} and \ref{Eq:mf_neqn_3a}.
Also, the MF-CLEAN formulation uses a two-term Taylor-polynomial, which can be shown to
result in a dynamic-range limit of $10^4$ for a bandwidth ratio of 2:1 and 
source spectral index of -1.0.

%



\subsection{Relation to MS-CLEAN}\label{Sec:relationMS}

The MS-CLEAN algorithm \citep{MSCLEAN, ERIC_MSCLEAN} models the sky as a combination
of multiscale flux components with no spectral structure.

A narrow-band (or flat-spectrum) version of MS-MFS can be derived by
setting $\Nt =1$ in the derivations in sections \ref{Sec:meqn} and \ref{Sec:neqn}.
%
\noindent The normal equations can be written in block matrix form (for example, for $\Ns =2$).
The peaks of the convolution kernels from the diagonal blocks of the Hessian
 are a measure of the sensitivity of the instrument
to a particular spatial scale. 
\begin{equation}
\left[\begin{array}{ll} 
\noalign{\medskip}
   [\He_{0,0}] & [\He_{0,1}] \\  
\noalign{\medskip}
   [\He_{1,0}] & [\He_{1,1}]  \\  
\noalign{\medskip}
   \end{array} \right]
\left[\begin{array}{l} \vec{I}^{\rm sky,\rm{\delta}}_{0} \\ 
\noalign{\medskip}
                       \vec{I}^{\rm sky,\rm{\delta}}_{1} \end{array}\right] =
\left[\begin{array}{l} \vec{I}^{\rm dirty}_{0}\\
\noalign{\medskip}
		       \vec{I}^{\rm dirty}_{1} \end{array}\right] 
\label{neqn_ms_math}
\end{equation}
%

In MS-MFS, a diagonal approximation of this Hessian is used to compute the principal solution.
This is equivalent to normalizing the residual images (RHS vectors) by the sum of weights for
each spatial scale, before searching for peaks.

In both existing forms of MS-CLEAN, this normalization is replaced by
a scale bias, an empirical term that de-emphasises large spatial scales.
The scale bias $b_s = 1-0.6~s/s_{\rm max}$  used by \citet{MSCLEAN}
(where $s_{\rm max}$ is the width of the largest scale basis function)
is a linear approximation of how the inverse of the area under each
scale function changes with scale size\footnote
{g
When $s/s_{\rm max} = 1.0$ the bias term is $1.0 - 0.6 = 0.4$ which is approximately equal
 to the inverse of the area under a Gaussian of unit peak and 
 width, given by ${1.0}/{\sqrt{2\pi}} = 0.398$.}.
The algorithm described by \citet{ERIC_MSCLEAN} uses 
$b_s \approx 1.0/s^{2x}$ where $x\in\{0.2,0.7\}$, to approximate
a normalization by the area under a Gaussian,
for the case when images are smoothed by applying a $uv$-taper 
that tends to unity for the zero spatial-frequency.
Both these normalization schemes are 
approximations of using a diagonal approximation of $[\He^{\rm peak}]$. 

Once we have this understanding, we can see that the full Hessian $[\He^{\rm peak}]$ 
(and not just a diagonal approximation)
can be inverted to get the normalization exactly right, giving 
an accurate estimate of the total flux at each scale. This becomes useful for 
sources that contain overlapping flux components of different spatial scales. 
However, this solution gives correct values only at the locations of the
centers of the flux-components, and introduces large errors in the PSF sidelobes.
Therefore, for reasons of stability, a diagonal approximation is still a more 
appropriate choice (demonstrated on simulated EVLA data). 

Another difference lies in the minor-cycle updates.
The update steps in MS-MFS and  the Cornwell MS-CLEAN evaluate the
full LHS of the normal equations to update the smoothed residual images and subtract out
flux components within the image domain. 
This allows each minor cycle iteration to search for flux components across all 
spatial scales. The update step in the Greisen MS-CLEAN ignores the 
cross-terms, and performs a full set of minor cycle iterations on one scale at a time.
A choice between all these methods will depend on trade-offs between the accuracy within each minor cycle iteration, 
the computational cost per step, and optimized global convergence
patterns to control the total number of iterations.


\section{Hybrids of Narrow-Band and Continuum Techniques}

The preceding sections discussed multi-frequency solutions that used data from all measured
frequencies together, to take advantage of the combined spatial-frequency coverage.
However, there are some situations where single-channel methods used in combination
with multi-frequency-synthesis (and no built-in spectral model) will be able to deliver
scientifically useful wide-band reconstructions at the continuum sensitivity.



%

The basic idea of a hybrid wide-band method is to combine 
the advantages of single-channel imaging (simplicity and 
non-dependence on any spectral model) with those of continuum imaging 
(deconvolution with full continuum sensitivity).
\begin{enumerate}
\item Deconvolve each channel separately upto the single-channel sensitivity limit $\sigma_{chan}$.
Only sources brighter than $\sigma_{chan}$ will be detected and deconvolved.
\item Remove the contribution of bright (spectrally varying) sources by subtracting out
visibilities predicted from the model image cube.
At this stage, the peak residual brightness is at the level of the single-channel
noise limit $\sigma_{chan}$. 
\item Perform MFS imaging (flat-spectrum assumption)
on the continuum residuals to extract flux that lies between $\sigma_{chan}$ and $\sigma_{cont}$.
According to \cite{MFCLEAN_CCW} (and as discussed later in section \ref{Sec:Errs}), 
errors due to this flat-spectrum assumption become visible only above a dynamic
range of $\sim$1000 (for $\alpha=-0.7$ and a 2:1 bandwidth ratio). 
Therefore, as long as the sensitivity improvement between a single-channel and
the full band is less than $\sim$1000, this second step will incur no errors even if the
remaining flux has spectral structure. 
This requirement translates to $N_{chan}<10^6$, which is usually satisfied\footnote
{Even if visibilities are measured at a very high frequency resolution, they can be
averaged across frequency ranges upto the bandwidth-smearing limit for the
desired image field-of-view. 
}.
\item Add model images from both steps, and restore the results. 
For unresolved sources, it may be appropriate to use a clean-beam fitted for
the highest frequency, but in general, to not bias spectral information, all channels
should be restored using a clean beam fitted to the PSF at the lowest frequency in the range.
\end{enumerate}
The advantages of this approach are its simplicity, and that it can handle wide-band
reconstructions with band-limited signals and spectral-lines.
The disadvantages are that the angular resolution of the images and spectral information 
will be restricted to that given by the lowest frequency (a factor of two worse than what is
possible for the 2:1 bandwidth available with the EVLA L-band).
Also, the single-frequency spatial-frequency coverage may not be sufficient to unambiguously
reconstruct all the spatial structure of interest at all frequencies, which in turn will
affect spectra measured from these single-channel images.
In some cases, additional constraints can be used. 
\cite{SPATIOSPECTRALMEM} describe spatio-spectral 
MEM, an entropy based method in which single-channel imaging is done
along with a smoothness constraint applied across frequency.

In general, single-channel methods can be used for wide-band imaging mainly to construct an image of the
continuum flux. Only if there is sufficient single-frequency $uv$-coverage to reconstruct an
accurate model of the source structure (for example, fields of isolated point sources),
may reliable spectral information also be derived from such an approach.



\section{MS-MFS Imaging results}\label{Sec:IMAGINGRESULTS}
This section contains three imaging examples that demonstrate the MS-MFS algorithm's
basic capabilities. Section \ref{Sec:Errs} discusses various sources of error and how they
manifest themselves, and section \ref{Sec:Feasibility} demonstrates the limits to which the algorithm
can be reasonably pushed.

\subsection{EVLA simulation}
\paragraph{Data :} Wide-band EVLA observations were simulated for a sky brightness distribution 
consisting of one point source with spectral index of $-$2.0 and 
two overlapping Gaussians with spectral indices of $-$1.0 and $+$1.0. 
The spectral
index across the resulting extended source varies smoothly between $-$1.0 and $+$1.0,
with a spectral turnover in the region of overlap of the two Gaussians, 
corresponding to a spectral curvature of approximately +0.5. 
Data were simulated for the EVLA in C-configuration between 1-2 GHz, for an
8-hour observation with measurements 5 minutes apart. 
The goal of this test was to assess the ability of the MS-MFS algorithm to reconstruct
both spatial and spectral information accurately enough for astrophysical use.

\begin{figure}[t!]
\epsfig{figure=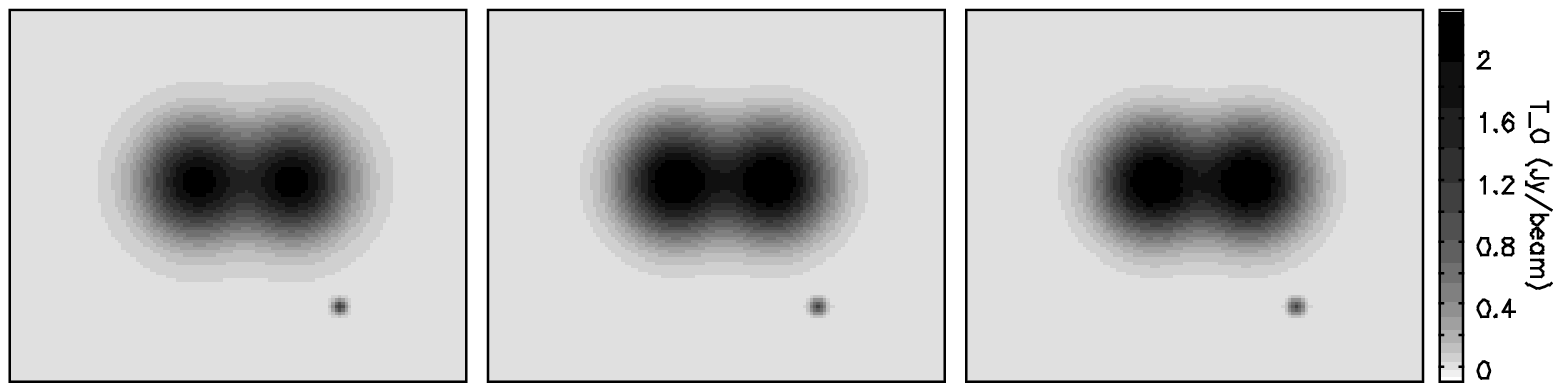,scale=0.565}
\epsfig{figure=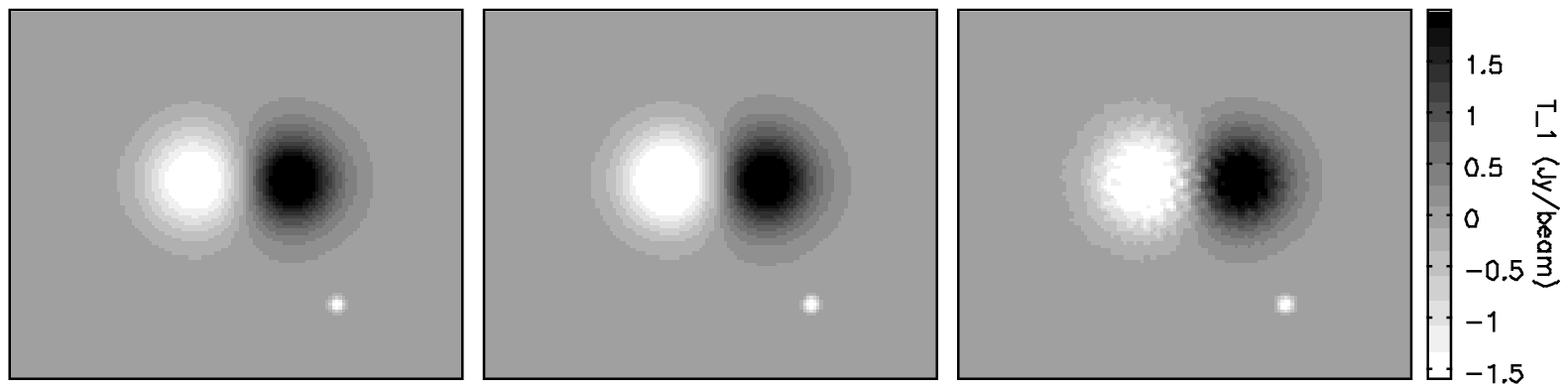,scale=0.565}
\epsfig{figure=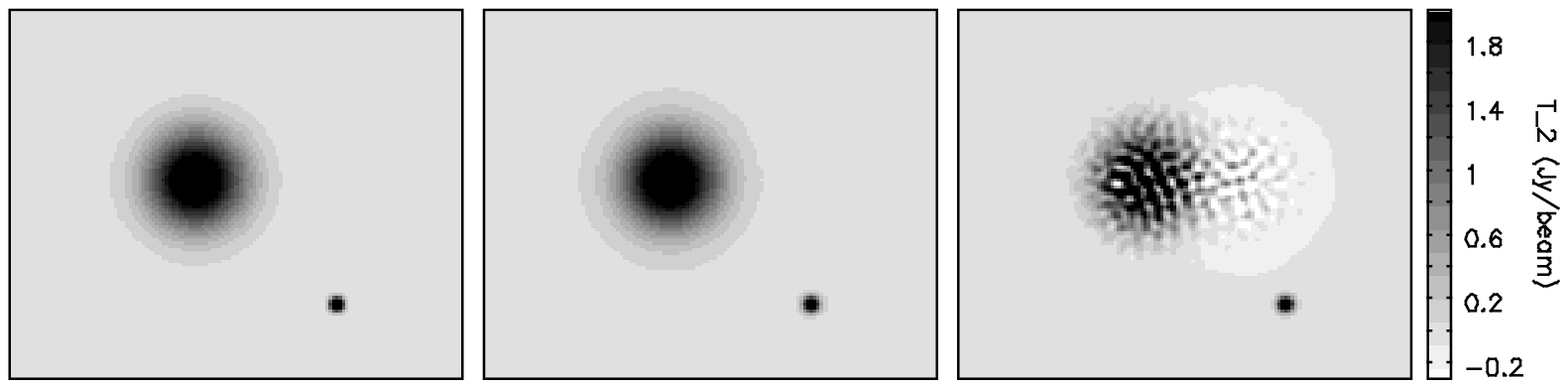,scale=0.565}
\epsfig{figure=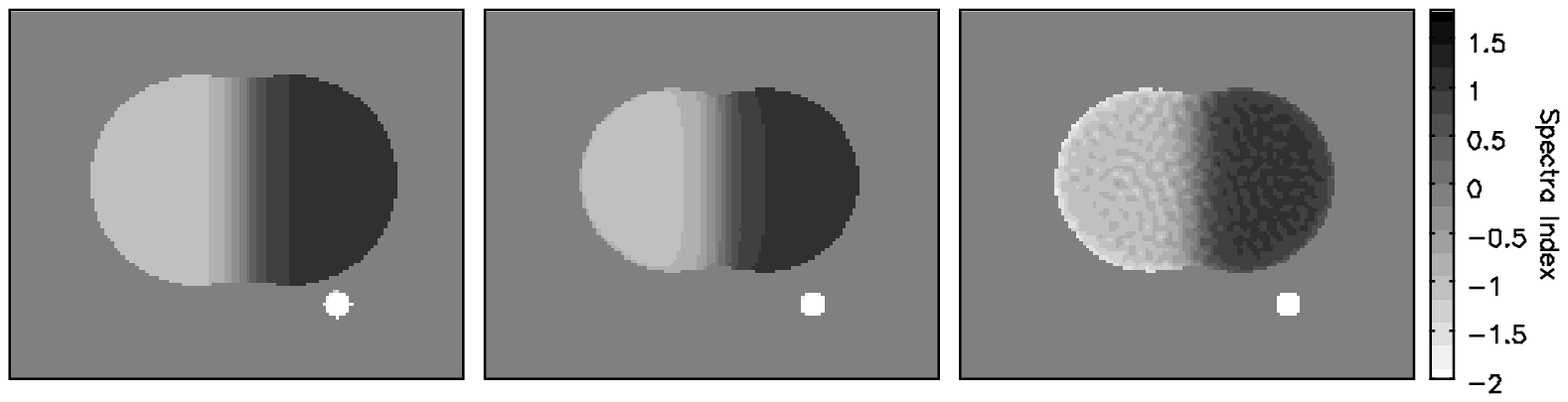,scale=0.565}
\epsfig{figure=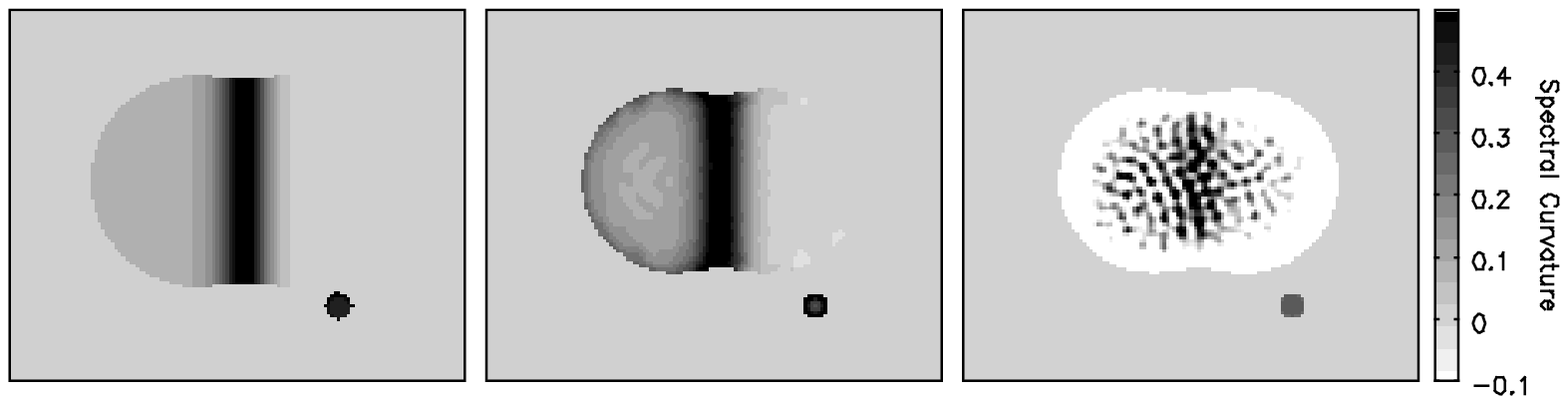,scale=0.565}
\caption[Example : Simulated wide-band sky brightness distribution]
{\small MS-MFS Imaging results using simulated EVLA data : 
These images compare truth images (left column) with the results of
two wide-band imaging runs; multi-scale (middle column) and point-source (right column).
The top three rows represent the first three Taylor-coefficient images,
and the fourth and fifth rows show spectral index and spectral curvature respectively.
}
\label{Fig:msmfs_sim}
\end{figure}

\paragraph{Imaging Results :} Two MS-MFS imaging runs were done and the results compared. 
Fig.\ref{Fig:msmfs_sim} illustrates the algorithm's performance by 
comparing several truth images describing the true sky brightness (left column) and 
reconstructions from a multi-scale (middle-column) and a point-source (right-column)
version of MS-MFS. The multi-scale version used a flux model
in which $\Nt =3$ and $\Ns =4$ with scale sizes defined by widths of $0,6,18,24$ pixels,
and the point-source version used $\Nt =3$ and $\Ns =1$ with one scale
function given by the $\delta$-function (to emulate the MF-CLEAN algorithm).
From top to bottom, the rows correspond to
the intensity image at the reference frequency $\I_{0} = \I_{\nuup_0}$,
the first-order Taylor-coefficient $\I_1 = \I_{\alphaup} \I_{\nuup_0}$, 
the second-order Taylor-coefficient $\I_2 = \left({\I_{\alphaup}(\I_{\alphaup}-1)}/{2} + \I_{\betaup} \right) \I_{\nuup_0}$,
the spectral index $\I_{\alphaup} = \I_1 / \I_0$ ,
and the spectral curvature $\I_{\betaup} = (\I_2/I_0) - \I_{\alphaup}(\I_{\alphaup}-1)/2$.

\begin{enumerate}
\item 
With a multi-scale multi-frequency flux model (MS-MFS) 
the spectral index across the extended source was reconstructed 
to an accuracy of $\delta\alpha < 0.02$ with the errors rising at the
edges of the source where the signal-to-noise ratio decreases.
The spectral curvature across the extended source was estimated to an accuracy of
$\delta\beta<0.05$ in the central region with the maximum error of 
$\delta\beta \approx 0.2$ in the regions 
where the curvature signal goes to zero and the source surface brightness
is also minimum (the outer edges of the source).

\item 
With a multi-frequency point-source model (MF-CLEAN)
the accuracy of the spectral index and curvature maps was limited to
$\delta\alpha\approx 0.2,\delta\beta\approx 0.6$.
This is because the use of a point source model will break any 
extended emission into components the size of the resolution element
and this leads to deconvolution errors well above the
off-source noise level.
Error propagation during the computation of spectral index and 
curvature as ratios of these noisy Taylor-coefficient images leads to
high error levels in the result (see section \ref{Sec:errprop}).
This clearly shows the importance of using a multiscale flux model when there
is extended emission.


\item The Gaussian on the left has $\alpha=-1.0$ and $\beta=0.0$, and $\Nt =3$ is not sufficient
to model the power-law accurately, leading to a value of $\beta\approx+0.1$ in the truth
image as well as in the reconstruction. However, the Gaussian on the right has $\alpha=+1.0$
and $\beta=0.0$ (a straight line), and $\Nt =3$ is more than sufficient to model it, leading to
an accurate value of $\beta \approx 0$ in the truth image and MS-MFS reconstruction. 
The point-source has $\alpha=-2.0$, and the error on $\beta$ is proportionally higher. 
The use of $\Nt >3$ reduces these errors.

\end{enumerate}


\subsection{Multi-frequency VLA observations of Cygnus-A}

\paragraph{Objective : }
Multi-frequency VLA observations of the bright radio galaxy Cygnus~A were used to
test the MS-MFS algorithm on real data and to test standard
calibration methods on wide-band data.
Cygnus~A is an extremely bright (1000 Jy) radio galaxy with 
a pair of bright compact hotspots ($\alpha=-0.5$) located
about 1 arcmin away from each other on either
side of a very compact flat-spectrum core, and extended radio lobes associated
with the hotspots with broad-band synchrotron emission at multiple spatial scales
($\alpha=-0.6 \sim -1.0$) \citep{CYGA_1996}.
The best existing images of Cygnus-A and its spectral structure
have been from large amounts of multi-configuration
narrow-band VLA data  \citep{CYGA_1991} designed to measure the spatial structure as
completely as possible at two widely separated frequencies (1.4 and 4.8 GHz).

The goal of this test was to use multi-frequency
snapshot observations of Cygnus~A to evaluate how well the MS-MFS algorithm is 
able to reconstruct both spatial and spectral structure
from measurements in which the single-frequency $uv$-coverage is insufficient
to accurately reconstruct all the spatial structure at that frequency.

Data were taken in April 2009 when the VLA was transitioning to the EVLA.
15 out of 27 antennas had new wideband EVLA feeds, but the correlator was that
of the VLA (narrow-band). 
Wideband data were taken as a series of narrow-band snapshots spread across 
8 hours and 9 distinct frequencies across the L-Band (30 min per frequency tuning). 
Flux calibration at each frequency was done {\it via} observations of 3C286.
Phase-only calibration was done using an existing narrow-band image of Cygnus~A at 1.4 GHz 
 \citep{CYGA_1991} as a model, and further self-calibration was done with the wide-band
flux model derived from the MS-MFS algorithm.
The final dataset used for imaging consisted of 9 spectral windows
each of a width of about 4 MHz and separated by about 100 MHz.

These data were imaged using two methods, the  
MS-MFS algorithm with $\Nt =3$ and $\Ns =10$, and a hybrid of single-channel imaging
followed by MFS on the residuals. Note that these observations do not have dense
single-frequency $uv$-coverage, and the purpose of applying the hybrid method was
to emphasize the errors that can occur if this method is used inappropriately.
Due to the small angular size of Cygnus-A, the effect of the L-band 
primary-beam was ignored in both runs.

\begin{figure}[t!]
\begin{center}
\epsfig{figure=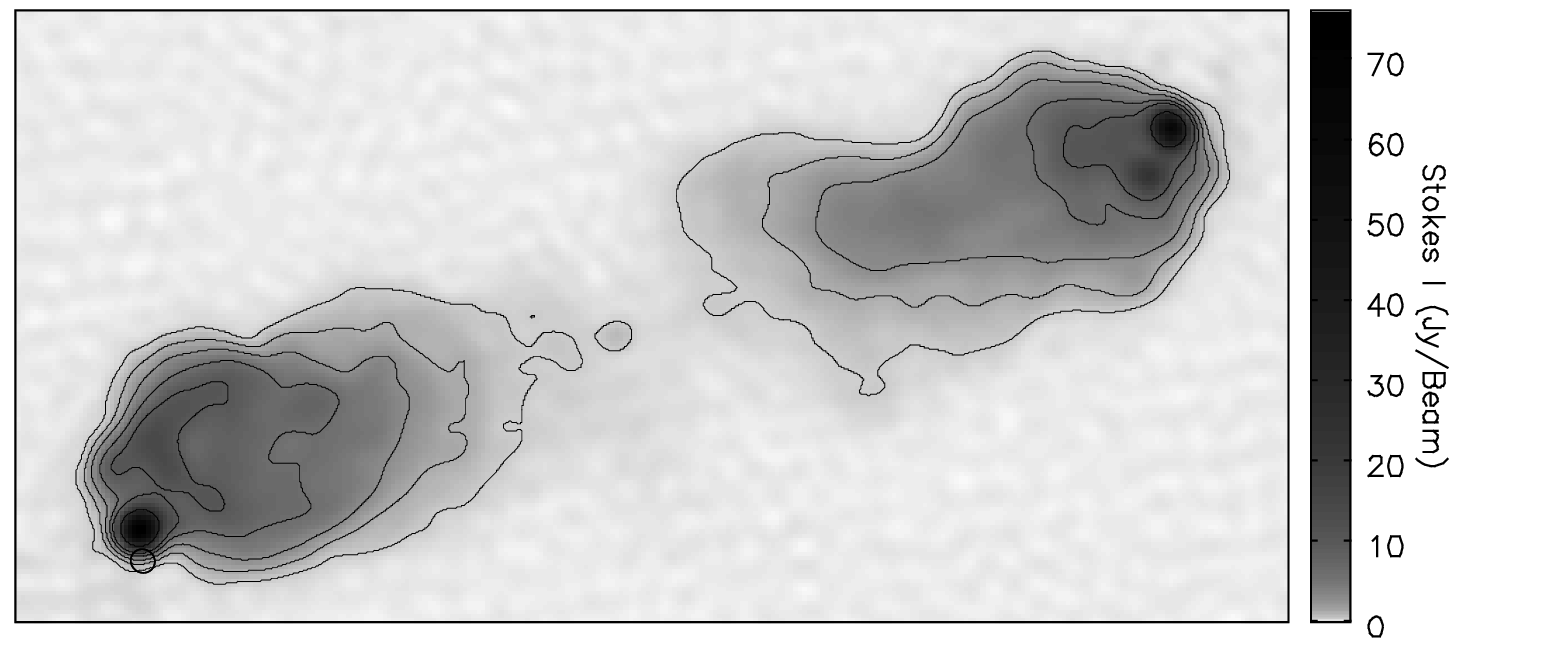,scale=0.45}
\epsfig{figure=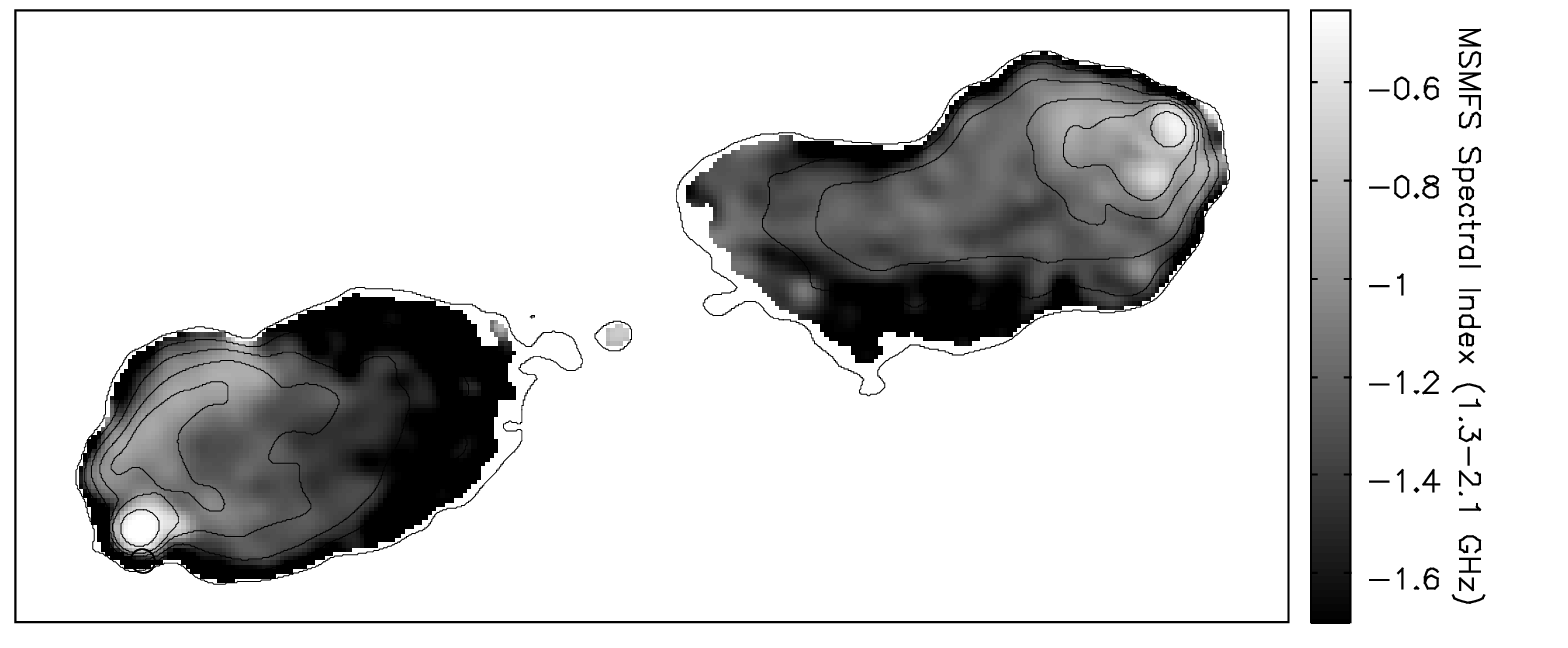,scale=0.45}
\epsfig{figure=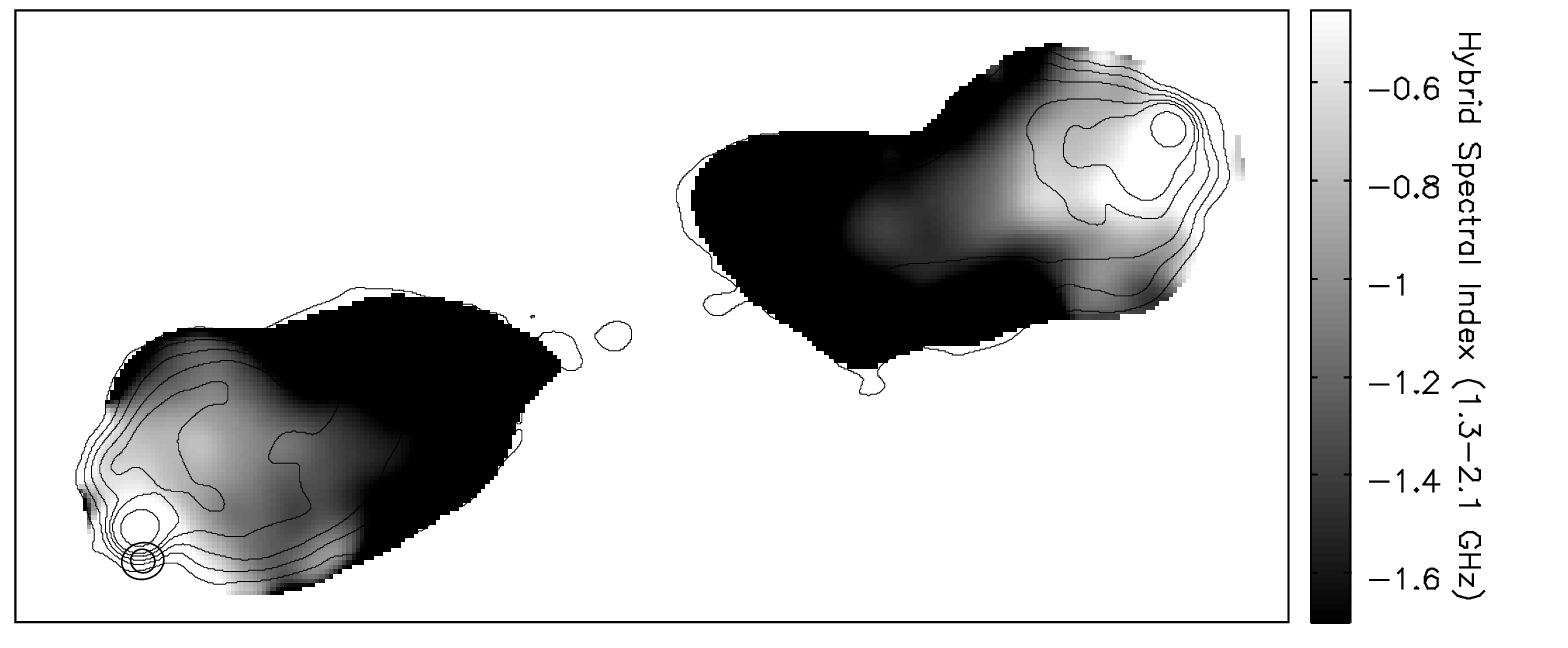,scale=0.45}
\epsfig{figure=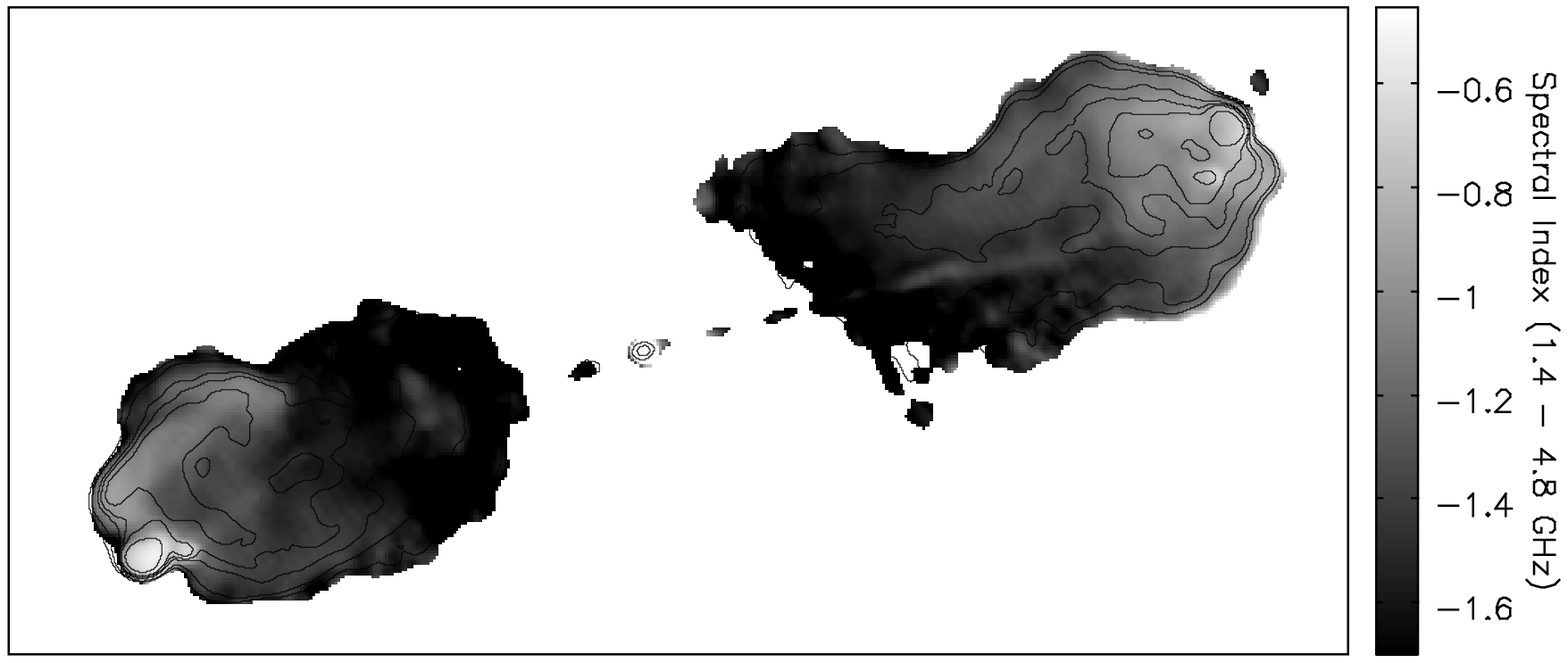,scale=0.45}
\end{center}

\caption[Cygnus~A : Intensity and residual images]
{\small Cygnus~A - total intensity and spectral-index :  
This figure shows the MS-MFS total intensity map (top), and spectral-index maps
obtained via three methods - MS-MFS (second row), a hybrid single-band method (third row), 
and from high-resolution full-synthesis narrow-band images
at 1.4 and 4.8 GHz (bottom). The MS-MFS spectral index map has the angular-resolution
of the total intensity map, and agrees with the values in the high-resolution comparison
map ($\alpha=-0.5$ at the hotspots increasing to $\alpha\approx-1.0$ in the halo). 
However, the hybrid method resulted in a map with a wider range of values 
(positive and negative) and is at a much lower angular resolution.
}
\label{Fig:cyga_intres}
\end{figure}

\paragraph{Results : }
Figure \ref{Fig:cyga_intres} shows the reconstructed total-intensity images (top), 
the spectral-index map obtained via MS-MFS (second row) and the spectral-index map
constructed from single-subband maps (third row).
For comparison, the image at the bottom is a spectral-index map constructed
from existing narrow-band images at 1.4 and 4.8 GHz, each constructed from 
a combination of VLA A, B, C and D configuration data \citep{CYGA_1991}.


\begin{enumerate}
\item The total intensity image (top row) 
has a peak brightness of 77 mJy/beam at the hotspots and a peak brightness of about
400 mJy/beam for the fainter extended parts of the halo. 
Both the methods (MS-MFS and hybrid) gave very similar total-intensity images
and residual images.
%
The on-source and off-source residuals for the MS-MFS algorithm are
30 mJy and 25 mJy and with the hybrid algorithm are 50 mJy and 30 mJy 
respectively.

\item The spatial structure seen in the MS-MFS spectral index image (second row) is very similar
to that seen in the two-point (1.4 - 4.8 GHz) spectral index image (bottom). 
This shows that despite having a comparatively small amount
of data (20 VLA B-configuration snapshots at 9 frequencies) the use of an algorithm that models
the sky brightness distribution appropriately is able to extract the same astrophysical
information as traditional methods applied to large amounts of data 
(full synthesis runs in multiple VLA configurations at two frequencies).
The estimated errors on the spectral index map are $<0.1$ for the brighter regions
of the source (near the hotspots) and $\ge 0.2$ for the fainter parts of the lobes
and the core.
In contrast, the Hybrid spectral index
map (third row) clearly shows errors arising due to non-unique 
solutions at each separate frequency (due to insufficient narrow-band 
spatial-frequency coverage) as well as smoothing to the
angular resolution at the lowest frequency.

\item The MS-MFS spectral curvature map (not shown here) contains values corresponding to
$\triangle\alpha\approx0.4$ for the brightest regions on the source. 
Such a change in $\alpha$ appears high, and we did not have sufficient
single-frequency data at other bands to verify these values (the next section
contains an example where we could verify the curvature maps with other data).
Also, the values of curvature changed between imaging-runs 
with $\Nt =3$ and $\Nt =4$, suggesting over-fitting errors due to the low signal-to-noise ratio of
the curvature measurement (which could arise from inaccurate bandpass calibration). 
\end{enumerate}

\subsection{Multi-frequency observations of M87}
A similar observation of M87 was done with the goal of measuring the 1-2 GHz spectral index
in different parts of the inner core-jet-lobe system and outer halo filaments.

M87 is a bright (200 Jy) radio galaxy located at the center of the Virgo cluster.
The spatial distribution of broad-band synchrotron emission from this source 
consists of a bright central region (spanning a few arcmin) containing a 
flat-spectrum core, a jet (with known spectral index of $-$0.55) and two radio lobes
with steeper spectra ($-$0.5 $> \alpha > -$0.8) 
\citep{Rottmann_1996_AA,Owen_2000}.
This central region is 
surrounded by a large diffuse radio
halo (7 to 14 arcmin) with many bright narrow filaments ($\approx 10''\times 3'$).

\begin{figure}[t!]
\begin{center}
\epsfig{figure=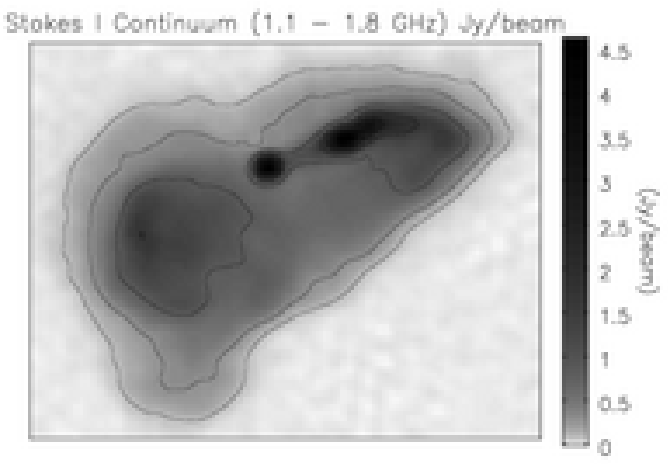,scale=0.4}
\epsfig{figure=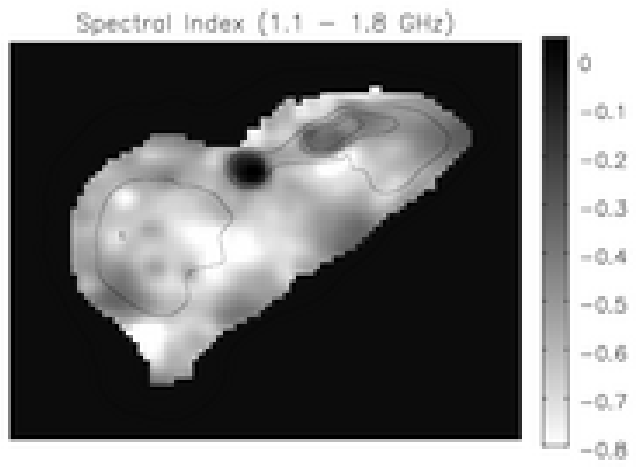,scale=0.4}
\epsfig{figure=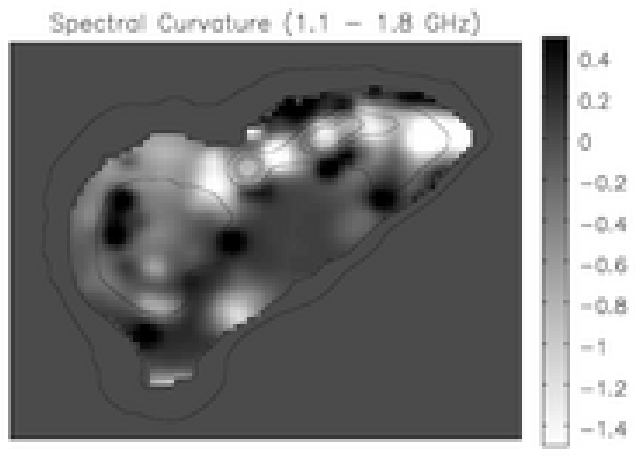,scale=0.4}
\end{center}
\begin{center}
\epsfig{figure=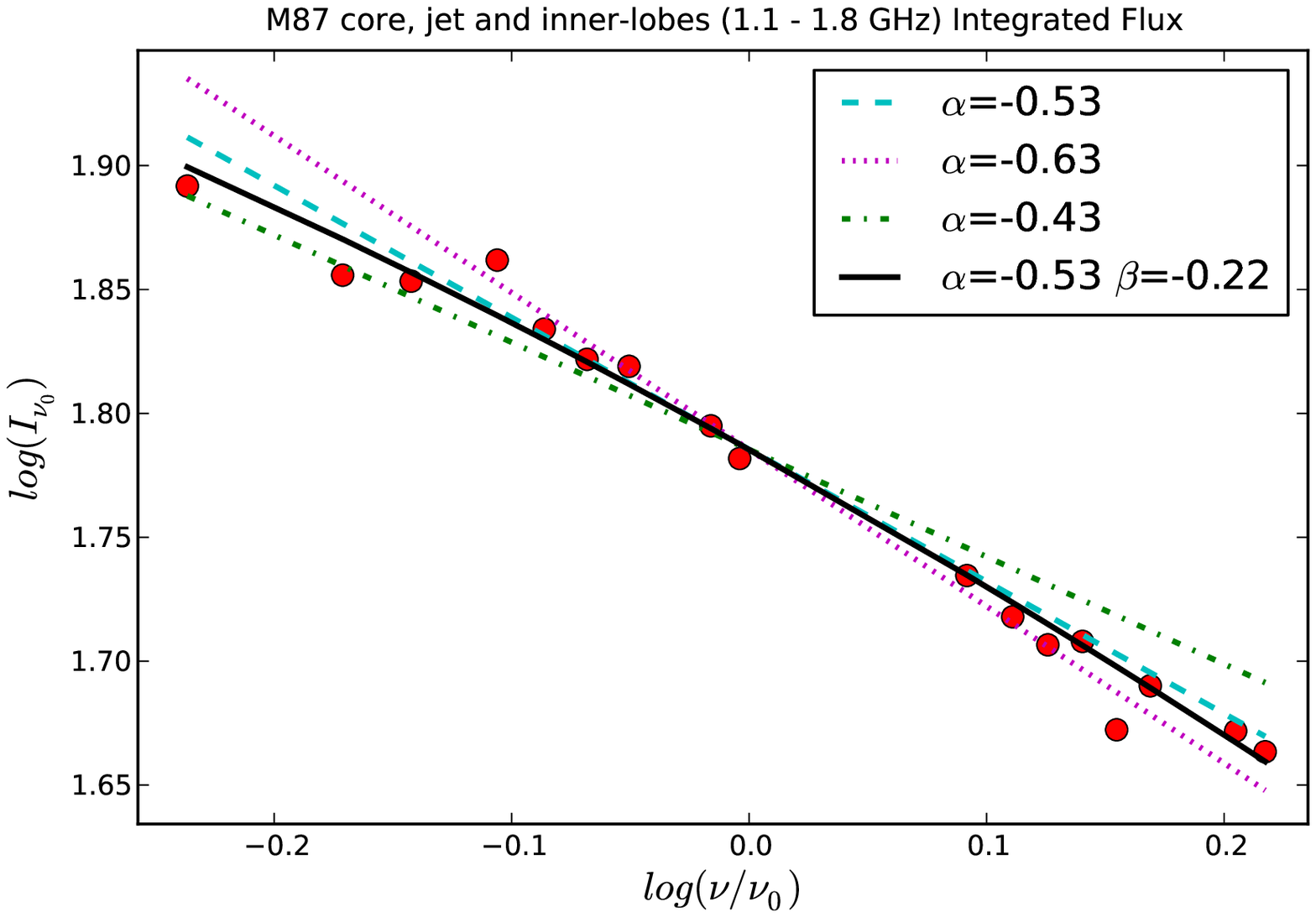,scale=0.5}
\end{center}
\caption[M87 core/jet/lobe - Intensity, Spectral index and Curvature]
{\small M87 core/jet/lobe - Intensity, Spectral index and Curvature : 
These images show 3-arcsec resolution maps of the central bright
region of M87 (core+jet and inner lobes).
The quantities displayed are the intensity at 1.5 GHz (top left),
the spectral index (top middle) and the spectral curvature (top right).
The spectral index is near zero at the core, varies between $-$0.36 and
$-$0.6 along the jet and out into the lobes. The spectral curvature is
on average 0.5 which translates to $\triangle\alpha=0.2$ across L-band.
The plot at the bottom compares the resulting average 
spectrum (solid line) with that formed by imaging each spectral-window separately (dots).
The dashed and dotted lines correspond to fixed values of spectral index (-0.43,-0.53,-0.63).
}
\label{Fig:m87_lobes}
\end{figure}

Multi-frequency VLA data were taken similar to the observations of 
Cygnus-A, with a series of 10 snapshots at 16 different frequencies 
within the sensitivity range of the EVLA L-band receivers. 
These data were imaged using MS-MFS with $\Nt =3$ and $N_2=12$.
The top row of images in 
Fig.\ref{Fig:m87_lobes} shows the intensity, spectral
index and spectral curvature maps of the bright central region at an
angular resolution of 3 arcsec (C+B-configuration).

\begin{enumerate}
\item 
The peak brightness at the center of the final restored intensity image
was 15 Jy with an off-source RMS of 1.8 mJy and an
on-source RMS of between 3 and 10 mJy.
%
The spectral index map\footnote
{The spectral index between two frequency bands $A$ and $B$ will be denoted 
as $\alpha_{\rm AB}$. For example, the symbol $\alpha_{\rm PL}$ corresponds to the
frequency range between P-band (327 MHz) and L-band (1.4 GHz), and
$\alpha_{\rm LL}$ corresponds to two frequencies within L-band (here, 1.1 and 1.8 GHz).
A similar convention will be used for spectral curvature $\beta$.
} of the bright central region (at 3 arcsec resolution)
shows a near flat-spectrum core with $\alpha_{\rm LL} = -0.25$,
a jet with $\alpha_{\rm LL} = -0.52$ and lobes with $-0.6 > \alpha_{\rm LL} > -0.7$.
This bright central region had sufficient
($>$100) signal-to-noise to be able to detect spectral curvature, and no
obvious deconvolution errors. 
However the error bars on the spectral curvature are at the same level as 
the measurement itself, and a reliable estimate can only be obtained as
an average over this entire bright region. The average curvature is
measured to be $\beta_{\rm LL} = -0.5$ which corresponds to a change in $\alpha$ 
across L-band by
$\triangle \alpha = \beta \frac{\triangle\nu}{\nu_0} \approx -0.2$.

\item To verify consistency of this spectral-curvature value 
with the data, each of the 16 spectral-windows was 
imaged separately, and restored with the same clean-beam.
The plot at the bottom of Fig.\ref{Fig:m87_lobes} shows this integrated flux spectrum 
($\log(I)$ vs $\log(\nu/\nu_0)$) as round dots, along with the average spectrum calculated
by MS-MFS (curved line), and straight dashed and dotted lines that correspond
to constant spectral indices of $-$0.43, $-$0.53 and $-$0.63. These plots show that a change in
$\alpha$ of about 0.2 is consistent with the data.
Note that the scatter seen in the single-spectral-window data points is at the
$1\%$ level of the source flux. This illustrates the accuracy at which bandpass
calibration must be done in order to measure a physically plausible spectral-curvature
signal across a 2:1 bandwidth.
%
%
\item These numbers were further compared with two-point spectral indices computed between
327 MHz (P-band), 1.4 GHz (L-band), and 4.8 GHz (C-band) from existing images
\citep{Owen_2000},(Owen, F. (private communication)). 
Across the bright central region, two-point spectral indices are $-0.36 > \alpha_{\rm PL} > -0.45$ and 
$-0.5 > \alpha_{\rm LC} > -0.7$. The measured values from our experiment ($-0.4 > \alpha_{\rm LL} > -0.7$
and $\triangle\alpha \approx 0.2$ across L-band) are consistent with these
independent calculations.


\end{enumerate}

\section{Sources of Error}\label{Sec:Errs}

There are various sources of error that can affect the imaging process, and leave artifacts
both on-source and off-source. As with any image-reconstruction algorithm, 
signs of these errors must be looked-for
in the output images before astrophysical interpretation.

\subsection{On-source polynomial-fit errors}

The errors on the polynomial coefficients and quantities derived from them
will depend on the number of measurements of the spectrum, 
their distribution across a frequency range,
the signal-to-noise ratio of the pattern being fitted, and 
the order of the polynomial used in the fit.
These errors will affect regions in the image both on-source and off-source, and
the resulting error patterns and their magnitudes will depend on the 
available $uv$-coverage, and the choice of reference frequency.
Although the physical parameters $\I_{\nuup_0},\I_{\alphaup}$ and $\I_{\betaup}$ can be obtained from 
the first three coefficients of a Taylor expansion of a power-law with varying index
(Eqns.\ref{Eq:calcab_1} to \ref{Eq:calcab_3}),
a higher order polynomial may be required during the fitting process to 
improve the accuracy of the first three coefficients\footnote
{\citet{MFCLEAN_CCW} comment on a bias that occurs with a 2-term Taylor expansion, due to the 
use of a polynomial of insufficient order to model an exponential.}.
%

\begin{figure}[t!]
\begin{center}
\epsfig{figure=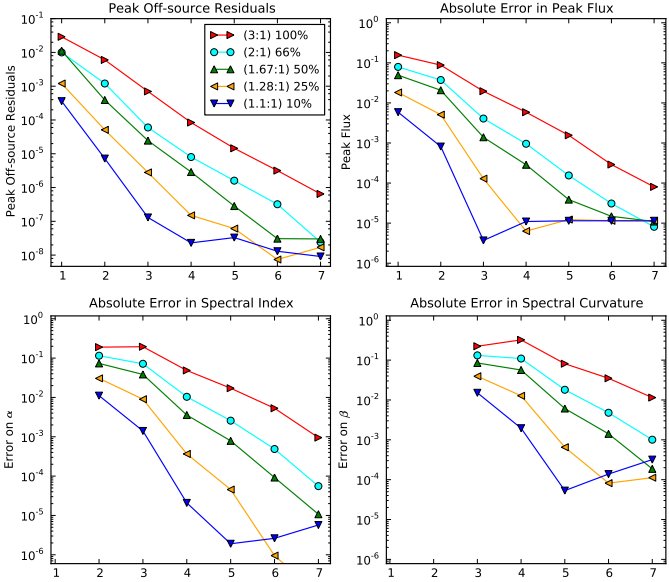,scale=0.44}
\end{center}
\caption[Peak residuals and errors for MFS with different values of $\Nt $]
{\small Peak Residuals and Errors for MFS with different values of $\Nt $ :
These plots show the measured peak residuals (top left) and the errors
on $\I_{\nuup_0}$(top right), $\I_{\alphaup}$ (bottom left), and $\I_{\betaup}$ (bottom right)
when a point-source
of flux 1.0 Jy and $\alpha$=-1.0 was imaged using Taylor polynomials of different
orders ($\Nt =1-7$) and a linear spectral basis. 
The x-axis of all these plots show the value of $\Nt $ used for the simulation
and plots for $\alpha$ and $\beta$ begin from $\Nt =2$ and $\Nt =3$ respectively.
%
An example of how to read these plots : For a 2:1 bandwidth ratio, 
a source with spectral index = -1.0 and $\Nt =4$,
the achievable dynamic range (measured as the ratio of the peak flux to the 
peak residual near the source) is about $10^5$, the error on the peak flux
at the reference frequency is 1 part in $10^3$, and the absolute
errors on $\alpha$ anre $\beta$ are $~10^{-2}$ and $~10^{-1}$ respectively.
}
\label{Fig:taylor_error}
\end{figure}


Figure.\ref{Fig:taylor_error} summarizes the errors obtained when the order of the polynomial
chosen for imaging is not sufficient to model the power-law spectrum of the
source. 
Data for 8 hour synthesis runa with EVLA $uv$-coverages 
were simulated for 5 different frequency ranges
around 2.0 GHz. The sky brightness distribution used for the simulation
was one point source whose flux is 1.0 Jy and spectral index is -1.0 with
no spectral curvature.
The bandwidth ratios\footnote
{There are two definitions of bandwidth ratio that are used in radio 
interferometry. One is the ratio of the highest to the lowest frequency
in the band, and is denoted as $\nu_{\rm high}:\nu_{\rm low}$. Another definition
is the ratio of the total bandwidth to the central frequency 
$(\nu_{\rm high}-\nu_{\rm low})/\nu_{\rm mid}$ expressed as a percentage.
For example, the bandwidth ratio for
$\nu_{\rm low}=1.0$ GHz, $\nu_{\rm high}=2.0$ GHz is $2:1$ and $66\%$.
}for these 5 datasets were
100\%(3:1), 66\%(2:1), 50\%(1.67:1), 25\%(1.28:1), 10\%(1.1:1).

MS-MFS was repeated on all these datasets with $\Nt =1$ to $\Nt =7$, 
and one scale $\Ns =1$, a $\delta$-function. 
All these datasets were imaged using a maximum of 10 iterations, a 
loop-gain of 1.0, natural weighting and a flux threshold of 1.0 $\muup$Jy. No noise
was added to these simulations (in order to isolate and measure 
numerical errors due to the spectral fits).
The top left panel in Fig.\ref{Fig:taylor_error} shows the peak residuals, measured over 
the entire $0^{th}$ order residual image. The other three panels show on-source
errors for $\I_{\nuup_0}$ (top right),$\I_{\alphaup}$ (bottom left) and $\I_{\betaup}$ (bottom right).
Errors on $\I_{\nuup_0},\I_{\alphaup},\I_{\betaup}$ were computed
at the location of the point source by taking differences with
the ideal values of $I_{\nuup_0}=1.0,\alpha=-1.0,\beta=0.0$.

One noticeable trend from these plots is that 
with sufficient signal-to-noise in the measurements, 
all errors decrease exponentially (linearly in log-space)
as a function of increasing order of the polynomial, and as a function
of decreasing total bandwidth. 
As expected, for very narrow bandwidths, the use of
high-order polynomials increases the error.
Also, when the order of the polynomial used is too low, 
the peak residuals are much smaller than the on-source error incurred on 
$\I_{\nuup_0}$, $\I_{\alphaup}$ and $\I_{\betaup}$. 
These trends are based on one simple example, and represent the
best-case scenario in which all sources can be described as point sources.
For extended emission, there can be additional errors due to deconvolution artifacts.
%
%
In the case of very noisy spectra, or at a late stage of image reconstruction when 
the signal-to-noise ratio in the residual images is low, errors can arise from attempting 
to use too many terms in the polynomial fit.

\subsection{Off-source errors and dynamic-range limits}

\begin{figure}[t!]
\begin{center}
\epsfig{figure=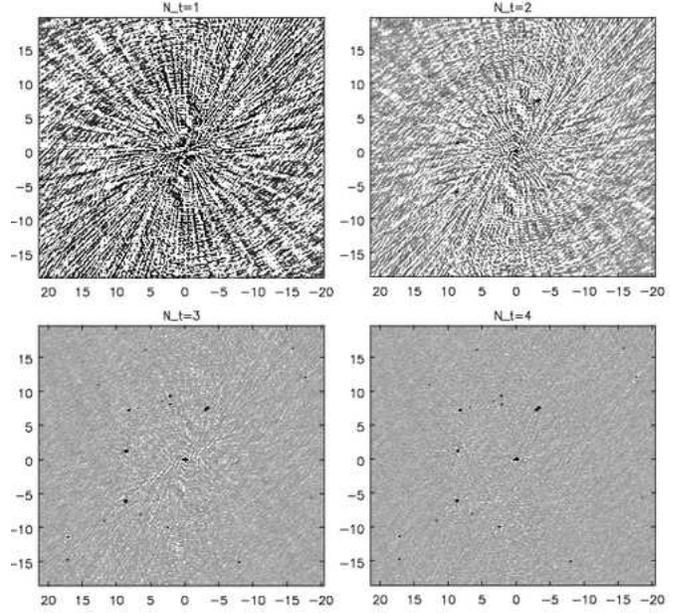,scale=0.5}
\end{center}
\caption{Stokes I images of the 3C286 field, using MSMFS with
$\Nt =1$ (top left), $\Nt =2$ (top right), $\Nt =3$ (bottom left), 
$\Nt =4$ (bottom left). The peak flux of 3C286 is 14 Jy/beam.
The axes labels are in units of arc-minutes from the pointing center
and all images are shown with the same grey-scales.
The residual RMS near 3C286 for these four images are
9 mJy,  1 mJy, 200 $\muup$Jy and 140 $\muup$Jy,  and the RMS
measured 1 degree away from 3C286 are 
1 mJy, 200 $\muup$Jy , 85 $\muup$Jy and 80 $\muup$Jy.
The thermal-noise limit for this dataset was 70 $\muup$Jy. 
}
\label{Fig:3C286}
\end{figure}

Consider the errors on the continuum image when spectral structure is ignored
during MFS imaging ($\Nt =0$; flat-spectrum assumption).
Spectral structure will masquerade as spurious spatial structure, leading to error patterns
that resemble the instrument's response to the first-order term in the Taylor-series expansion.
A rough rule of thumb for EVLA $uv$-coverages is that for a point source
with spectral index $\alpha=-1.0$ measured between 1 and 2 GHz,
the peak error  obtained if the spectrum is ignored is at a dynamic range of $<10^3$. 
In other words, if the dynamic-range allowed by the data (peak brightness / thermal noise)
is 800 (for example), there will be no visible artifacts if the spectral structure upto $\alpha=-1.0$ 
is ignored. 
These numbers are consistent with error estimates derived in  \cite{MFCLEAN_CCW} that 
predict errors at the level of $I\alpha/X$ where $X=O(500)$
is a factor that depends on the $uv$-coverage of the instrument and the choice of reference
frequency.  
Therefore, if the only goal is to obtain a continuum image over a narrow field-of-view, 
it may be possible to achieve the maximum-possible dynamic range 
by dividing out an average spectral index
(one single number over the entire sky) from the visibilities before imaging them
via MFS with a flat-spectrum assumption \citep{MFCLEAN_CCW}.


At high dynamic-ranges, off-source errors trace the spectral response functions for 
higher-order Taylor terms. As long as they are visible above the noise, they 
can be eliminated with a higher-order polynomial fit.
Figure \ref{Fig:3C286} shows a set of four images of the 3C286 field made from 
wideband EVLA data taken in October 2010. 
These data consist of
four EVLA snapshots spread across 90 minutes, and cover a frequency range
of 1.02 GHz to 2.1 GHz (800 MHz usable bandwidth after accounting for radio-frequency-interference). 
3C286 is 14.4 Jy/beam at 1.5 GHz, with a spectral index of -0.47. 
The four panels correspond to MS-MFS imaging
runs with $\Nt =1$ (top left), $\Nt =2$ (top right), $\Nt =3$ (bottom left), 
$\Nt =4$ (bottom left), all with $\Ns =1$ (a $\delta$-function).
All images are displayed with the same grey-scale levels, and show the
levels at which the error patterns appear. 
The dynamic range (calculated as the peak brightness to the peak residual
measured near the brightest peak) ranges from $1.6\times 10^{3}$ when spectral structure
is ignored ($\Nt =1$), to $1.1\times10^5$ with $\Nt =4$.


\subsection{Propagation of multi-scale errors}\label{Sec:errprop}

%
Deconvolution errors contribute to the on-source error in the Taylor 
coefficient images, and these errors propagate to the spectral index map which  
is computed as a ratio of two coefficient images ($\I_{\alphaup} = \I_1/\I_0$).
Errors that
result when a point-source flux model is used for extended emission
can increase the error bars on the spectral index and curvature by an order of magnitude
(as demonstrated by the example shown in Fig. \ref{Fig:msmfs_sim}).
These errors are approximately given by 

\begin{equation}
\triangle \I_{\alphaup} = \I_{\alphaup} \sqrt{ \left( {\triangle \I_0}\over{\I_0} \right)^2 + \left( {\triangle \I_1}\over{\I_1} \right)^2 }
\label{Eq:errprop}
\end{equation}
where $\triangle \I_0$ and $\triangle \I_1$ are the absolute errors measured in the first two
Taylor-coefficient images. 
For the example shown in Fig. \ref{Fig:msmfs_sim}, the errors on the coefficient images 
were measured as the RMS of the deviation from the truth images within the region of
high signal-to-noise. These errors result in a prediction of $\triangle \I_{\alphaup} \approx 0.05$ 
for the multi-scale version, and $\triangle \I_{\alphaup} \approx 0.7$ for the point-source version,
which approximately matches the RMS of the deviation of the output $\alpha$ image
from the truth $\alpha$ image.

\subsection{Frequency dependence of the Primary Beam}

When imaging wide fields-of-view, sources away from the
pointing center will be attenuated by the value of the primary beam at each
frequency. 
The EVLA primary-beams across a 2:1 bandwidth contribute an extra spectral-index
of -1.4 at the half-power point (measured from simulated beams, as well as a
Gaussian approximation of the main lobe of the primary beam \citep{MFCLEAN}).
Therefore, even if a source has a flat spectrum, this artificial spectral index
can result in imaging artifacts at the levels described in section \ref{Sec:Errs}
(i.e. a dynamic range limit of $\sim 10^3$ for a flat spectrum source at 
the HPBW, due only to the spectral variation of the primary beam).
This effect can be corrected as a post-deconvolution step, or by using wide-field
 imaging algorithms along with MS-MFS. So far, accurate post-deconvolution
corrections have been demonstrated 
out to the 10\% level of the highest-frequency primary-beam.

\section{Imaging Performance (non-standard conditions)}\label{Sec:Feasibility}
This section describes a set of simulations that test the limits of MS-MFS for different types
of source structure.  Three cases are studied; sources unresolved at the lowest
sampled frequency and resolved at the upper end, sources whose visibility functions 
lie mostly within the central unsampled region of the $uv$-plane near the origin, and
band-limited signals.

\subsection{Moderately resolved sources}

Consider a source with broad-band continuum emission and  
spatial structure that is unresolved at the low-frequency end of the band and resolved
at the high-frequency end.
Traditionally, spectral information would be available only at the angular 
resolution offered by the lowest observed frequency.
With MS-MFS, the intensity distribution as well as
the spectral index of such emission can be imaged
at the angular resolution allowed by the highest frequency in the band.
This is because compact emission has a signature all across the spatial-frequency plane
and its spectrum is well sampled by the measurements. The highest frequencies constrain
the spatial structure, and the flux model (in which a spectrum is associated with each
flux component) naturally fits a spectrum at the angular resolution at which the spatial 
structure is modeled. 
Such a reconstruction is model-dependent and will be accurate only
if the spectrum at the smallest measured scales can really be represented as a polynomial.

\begin{figure}[t!]
\begin{center}
\epsfig{figure=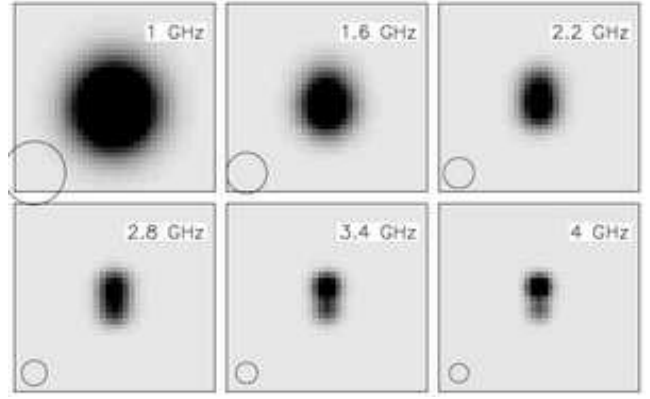,scale=0.8}
\end{center}
\caption[Moderately Resolved Sources : Single-Channel Images]
{\small Moderately Resolved Sources -- Single-Channel images : 
These figures show the 6 single-channel images generated from simulated
EVLA data between 1 and 4 GHz in the EVLA D-configuration. 
The sky brightness consists of two point sources, each of flux 1.0 Jy at
a reference frequency of 2.5 GHz and separated by 18 arcsec.
Their spectral indices are $+$1.0 (top source)  
and $-$1.0 (bottom source).
The angular resolution at 1 GHz is 60 arcsec, and at 4 GHz is 15 arcsec
and the circles in the lower left corner of each image shows the resolution
element decreasing in size as frequency increases.
}
\label{Fig:modres_channel}
\end{figure}

\begin{figure}[t!]
\begin{center}
\epsfig{figure=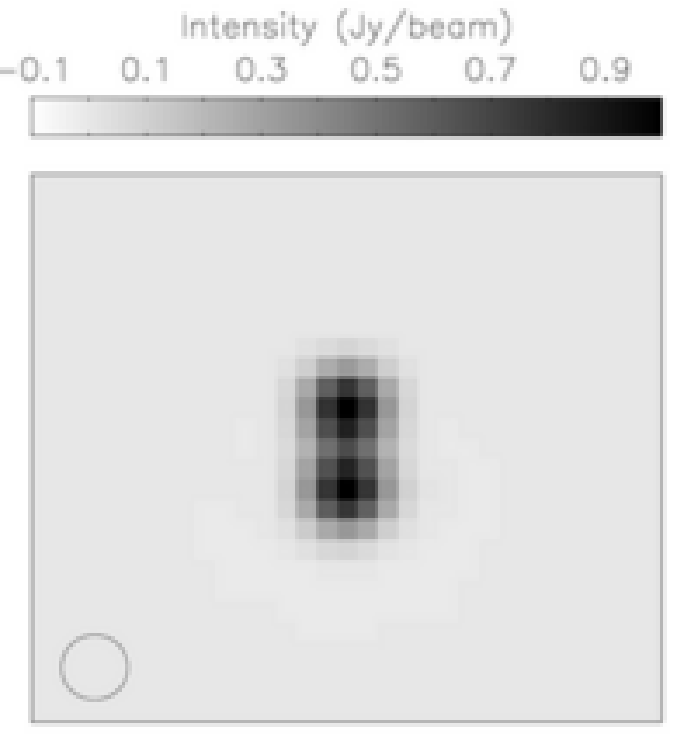,scale=0.4}
\epsfig{figure=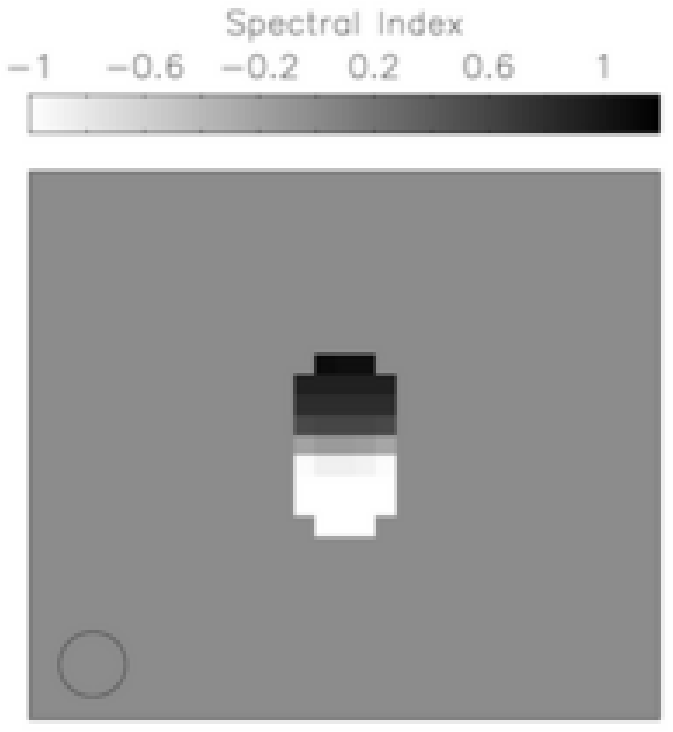,scale=0.4}
\epsfig{figure=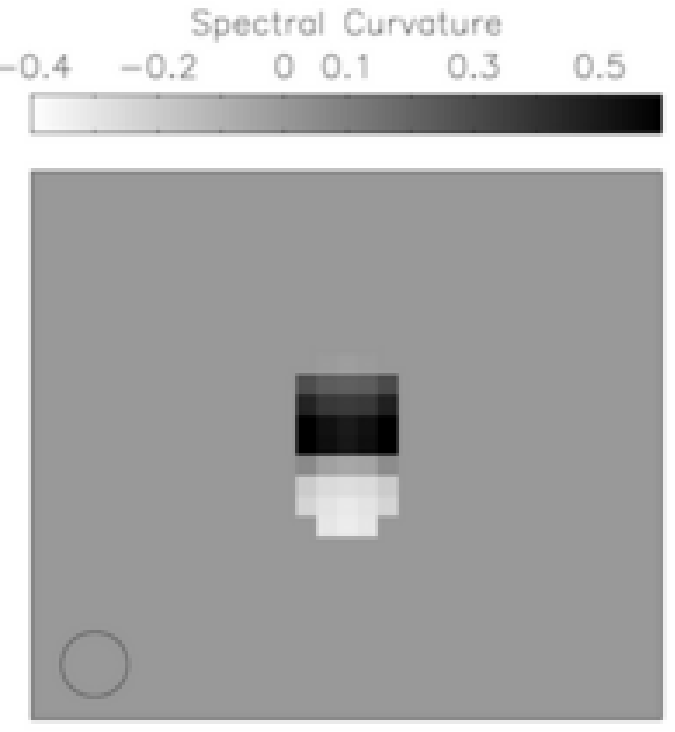,scale=0.4}
\end{center}
\caption[Moderately Resolved Sources : MSMFS Images]
{\small Moderately Resolved Sources -- MSMFS Images : 
These images show the intensity at 2.5 GHz (left),
the spectral index showing a gradient between $-$1 and +1 (middle)
and the spectral curvature which peaks between the
two sources and falls off on either side (right). 
The angular resolution of these images is 15 arcsec, corresponding to
the highest frequency in the data (and Fig.\ref{Fig:modres_channel}).
}
\label{Fig:modres_msmfs}
\end{figure}

\paragraph{EVLA Simulation : }
Wide-band EVLA data were simulated for the D-configuration
across a frequency range of 3.0 GHz 
with 6 frequency channels between 1 and 4 GHz (600 MHz apart).
This wide
frequency range was chosen to emphasize the difference in angular resolution
at the two ends of the band (60 arcsec at 1 GHz, and 15 arcsec at 4.0 GHz).
The sky brightness chosen for this test consists of a pair of
point sources separated by a distance of 18 arcsec (about one resolution
element at the highest frequency), making this a moderately resolved
source. These point sources were given different spectral indices
($+$1.0 for the top source and $-$1.0 for the bottom one).
Figure \ref{Fig:modres_channel} shows the six single-channel images of this
source.
At the low frequency end, the source is almost indistinguishable 
from a single flux component centered at the location of the bottom source
whose flux peaks at the low-frequency end. The source structure becomes
apparent only in the higher frequencies where the top source (with a positive
spectral index) is brighter.

\paragraph{MSMFS Imaging results : }
These data were imaged using the MS-MFS algorithm with $\Nt =3$ and 
$\Ns =1$ with only one spatial scale (a $\delta$-function), and
figure \ref{Fig:modres_msmfs} shows the results.
The intensity distribution, spectral index and curvature of this source were
recovered at the angular resolution allowed by the 4.0 GHz samples
(15 arcsec). 
This demonstrates that an appropriate flux model can constrain the 
solution accurately at the angular resolution given by the highest sampled
frequency.

\subsection{Emission at large spatial scales}

\begin{figure}[t!]
\begin{center}
\epsfig{figure=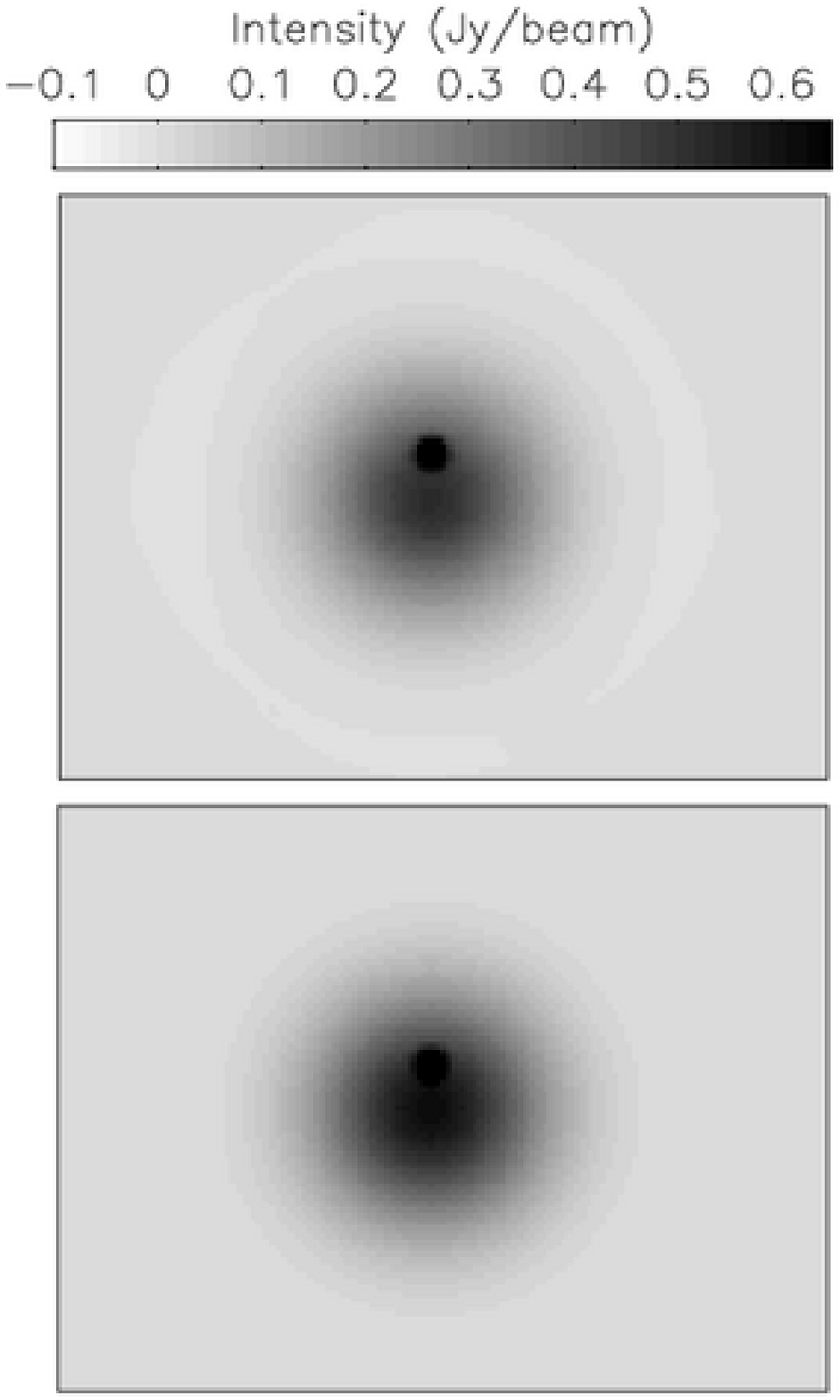,scale=0.25}
\epsfig{figure=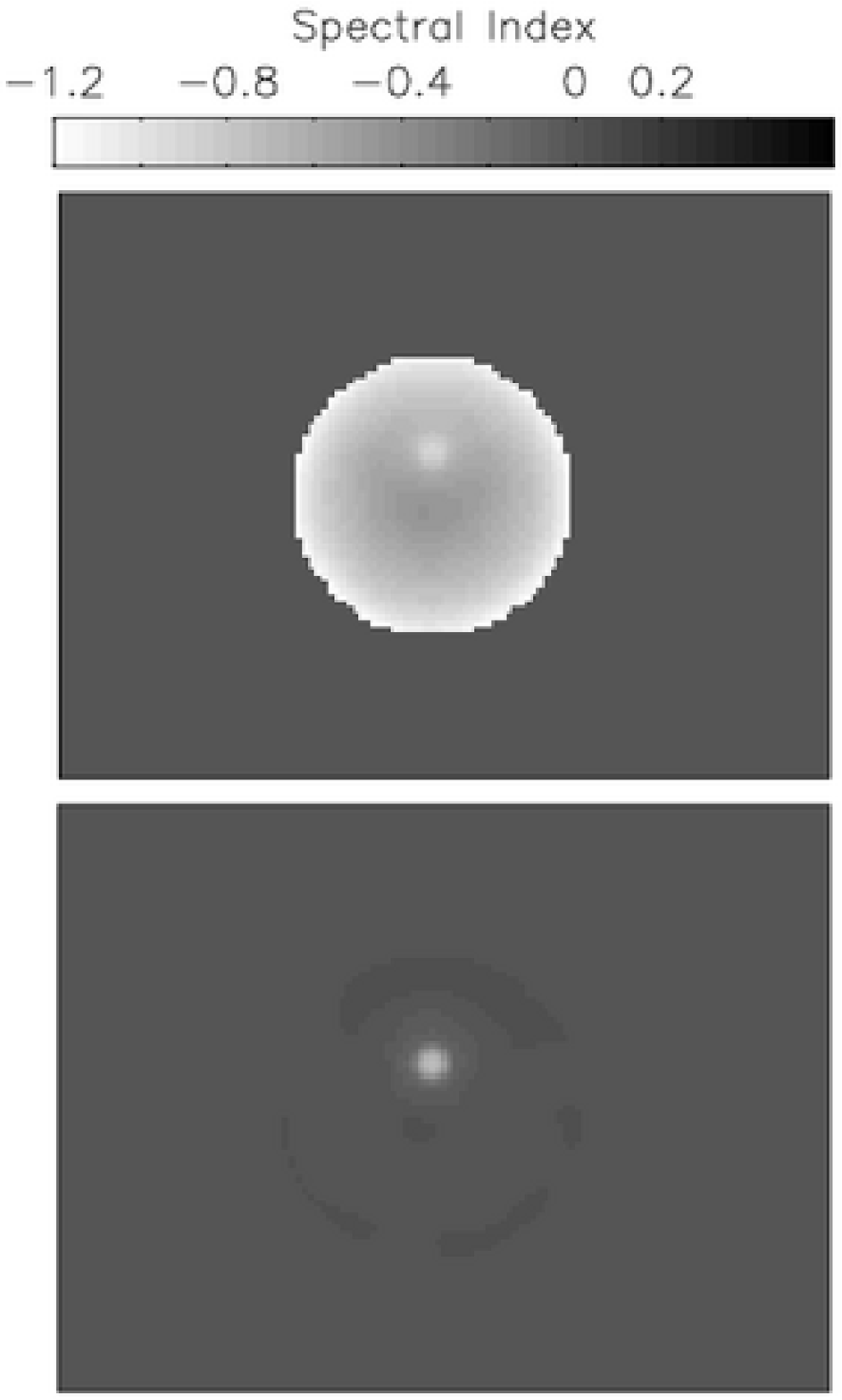,scale=0.25}
\epsfig{figure=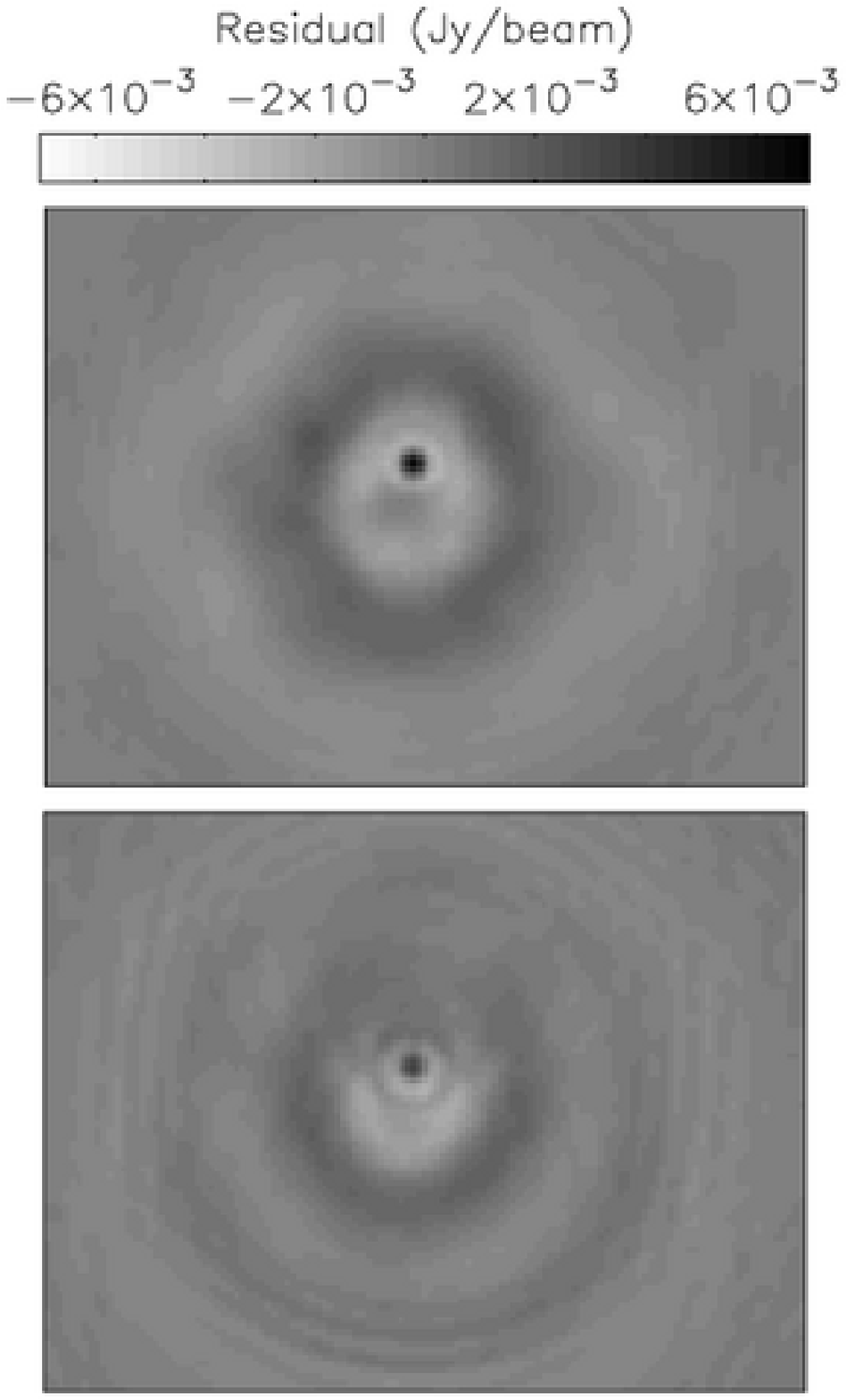,scale=0.25}
\end{center}
\caption[Very Large Spatial Scales : Intensity, Spectral Index, Residuals ]
{\small Very Large Spatial Scales - Intensity, Spectral Index, Residuals : 
These images show the intensity distribution (left), spectral index
(middle) and the residuals (right) for two imaging runs.
The top row shows results without short-spacing information, and shows a false
negative $\alpha\approx-0.8$ for the extended emission. The bottom row
shows results with short-spacing constraints, in which the extended source
has been reconstructed correctly.
}
\label{Fig:zero_images}
\end{figure}

Consider a very large (extended) flat-spectrum source whose  
visibility function falls mainly within the
central hole in the $uv$-coverage.
The size of this central hole increases with observing frequency.
The minimum spatial-frequency sampled per 
channel will measure a decreasing peak flux level as frequency increases.
%
Since the reconstruction below the minimum spatial-frequency
involves an extrapolation of the measurements 
and is un-constrained by
the data, these decreasing peak visibility levels can be mistakenly interpreted as
a source whose amplitude itself is decreasing with frequency
(a less-extended source with a steep spectrum).
Usually, a physically-realistic flux model suffices as a constraint, but with the 
MS-MFS model (polynomial spectra associated with 2D Gaussian-like components),
a large flat-spectrum source and a smaller steep-spectrum source are both
allowed and considered equally probable.
This creates an ambiguity between the reconstructed scale and spectrum that 
cannot always be resolved directly from the data, and 
requires additional information (a low-frequency narrow-band image
at the reference frequency to constrain the spatial structure, 
or low-resolution spectral information).

\begin{figure}[t!]
\begin{center}
\epsfig{figure=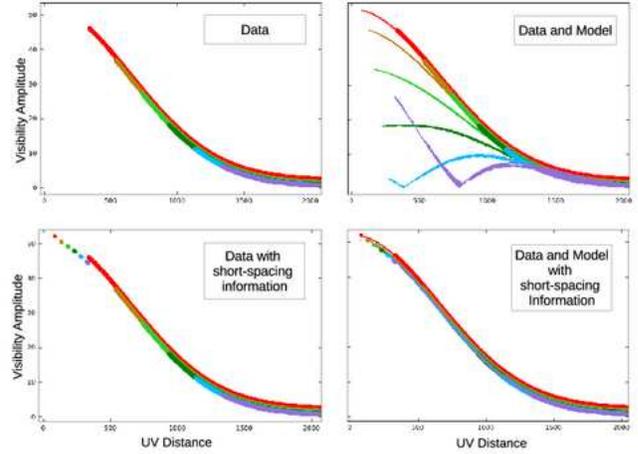,scale=0.5}
\end{center}
\caption[Very Large Spatial Scales : Visibility plots]
{\small Very Large Spatial Scales - Visibility plots : 
These plots show the observed (left) and 
 reconstructed (right) visibility functions for a
simulation in which a large extended flat-spectrum source is observed
with an interferometer with a large central hole in its $uv$-coverage.
The colours/shades in these plots represent
6 frequency channels spread between 1 and 4 GHz.
The top row of plots shows that when no short-spacing information was used, 
 these data can be mistakenly fit using a less-extended source with a steep
spectrum.
The bottom row shows that the inclusion of short-spacing information is 
sufficient to reconstruct the sky brightness distribution correctly.
}
\label{Fig:zero_visplot}
\end{figure}

\paragraph{EVLA Simulation : }
Wide-band EVLA data were simulated for the D-configuration
across a frequency range of 3.0 GHz centred at 2.5 GHz.
(6 frequency channels located 600 MHz apart between 1.0 and 4.0 GHz).
The size of the central hole in the $uv$-coverage was increased
by flagging all baselines shorter than 100 m 
and the wide frequency range was chosen to emphasize the difference
between the largest spatial scale measured at each frequency
(0.3~k$\lambda$ or 10.3 arcmin at 1.0 GHz, 
 and 1.3~k$\lambda$ or 2.5 arcmin at 4.0 GHz).
The sky brightness chosen for this test consists of one large
flat-spectrum ($\alpha=0.0$) 2D Gaussian whose FWHM is 2.0 arcmin (corresponding
to 1.6~k$\lambda$, at the reference frequency of 2.5 GHz), and one
steep spectrum point-source ($\alpha$=-1.0) located on top of
this extended source at 30 arcsec away from its peak.

\paragraph{MS-MFS Imaging Results : }
These data were imaged using the MS-MFS algorithm with $\Nt =3$ and
$\Ns =3$ with scale sizes given by [0,10,30] pixels. 
Two imaging runs were performed with these parameters, without and
with short-spacing information.
Fig.~\ref{Fig:zero_images} shows images of the intensity (left), spectral
index (middle) and residuals (right) for these two runs, and 
Fig.~\ref{Fig:zero_visplot} shows the visibility amplitudes 
present in the simulated data (left column) as well as in the reconstructed model 
(right column) at each of the 6 frequencies. 
In both figures, the top and bottom rows 
correspond to runs without and with short-spacing information respectively
and each pair (in Fig.\ref{Fig:zero_images}) are displayed at the same intensity scale.

\begin{enumerate}
\item 
The first imaging run used only baselines longer than 100m
to simulate a large central hole in the $uv$-coverage. 
The spectrum of the point-source was correctly estimated as $-$1.0,
but the extended source acquired a false steep-spectrum ($\alpha\approx -0.8$).
The algorithm was able to reconstruct the correct flux and spectrum
of the extended source only if the multi-scale basis functions were carefully
chosen to match the known scale size (i.e. a stronger {\it a-priori} constraint).
This shows that without additional constraints, it is not always possible
to distinguish large flat-spectrum source from a slightly less-extended steep spectrum source.
\item 
A second imaging run was performed including
additional information in the form of short-spacing 
constraints, 
added in by retaining a small number of very short-baseline 
measurements at each frequency 
(baselines shorter than 25 m).
%
%
The visibility plots and imaging results with this dataset 
show that the short-spacing flux estimates were sufficient to
bias the solution towards the correct solution (flat-spectrum extended source).  
\item The image residuals are at the same
level in both runs, demonstrating that in the absence of  
additional short-spacing information, both flux
models are equally poorly constrained by the data themselves.
One way to avoid this problem altogether (but lose some information)
is to flag all spatial-frequencies smaller than $u_{\rm min}$ at $\nu_{\rm max}$
and not attempt to reconstruct any spatial scales larger than what
$\nu_{\rm max}$ allows. 
\end{enumerate}

\subsection{Band-limited signals}
Consider a source of radiation where frequency traces different physical structures 
in the source (as opposed to a fixed structure with each point emitting broad-band radiation).
For such sources, emission may be detected 
in only part of the sampled frequency range and is for all practical purposes
a band-limited signal.

Since the MS-MFS algorithm uses a polynomial  to model the spectrum
of the source (and is not restricted to a power-law spectrum) it
should be able to reconstruct such structure as long as it varies smoothly.
However, for a band-limited signal, the angular resolution 
at which the structure can be mapped will be limited to the resolution of the 
highest frequency at which the signal is detected (and not the highest resolution
allowed by the measurements).

EVLA D-configuration data were simulated for the band-limited radiation
observed with synchrotron emission from solar prominences where different
frequencies probe different depths in the solar atmosphere. The structures are generally
arch-like with lower frequencies sampling the top of the loop and higher frequencies
sampling the legs.
The MS-MFS algorithm was run on these simulated data, using $\Nt =5$ to
fit a $4^{th}$-order polynomial to the source spectrum (to accomodate its
nearly band-limited nature) 
and $\Ns =3$ with scales given by [0, 10, 30] pixels.  
Iterations were terminated after 200 iterations.

\begin{figure}[t!]
\begin{center}
\epsfig{figure=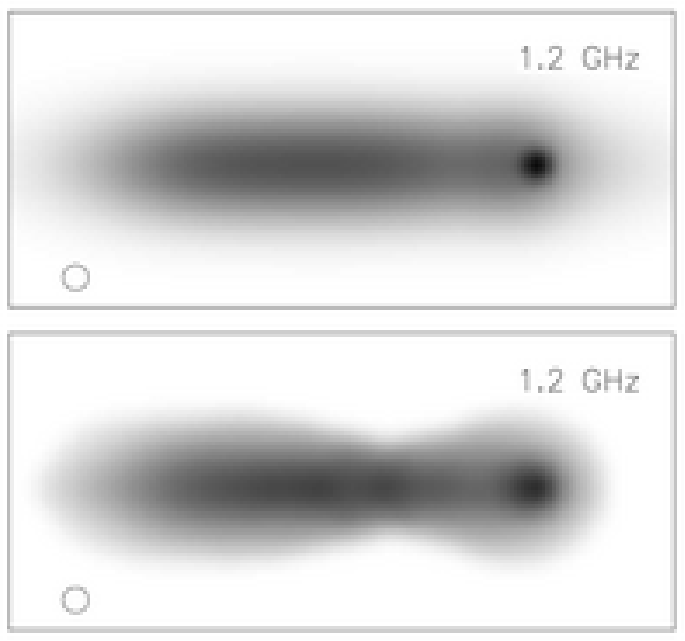,scale=0.4}
\epsfig{figure=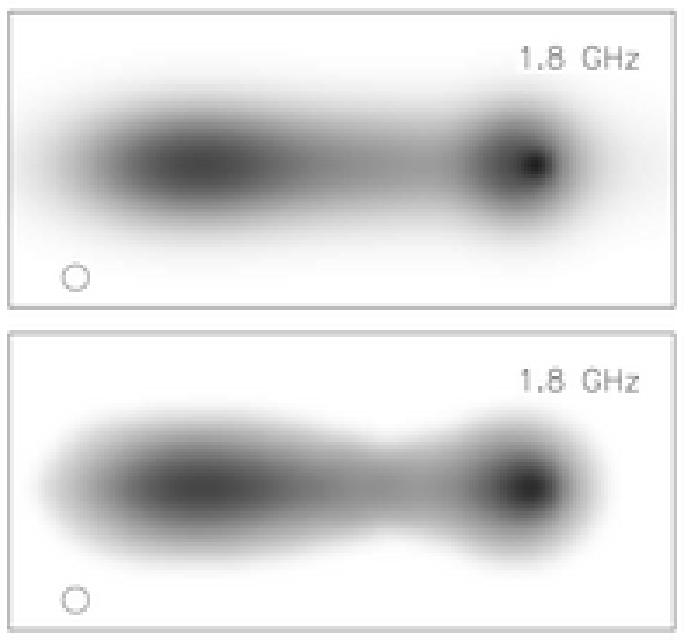,scale=0.4}
\epsfig{figure=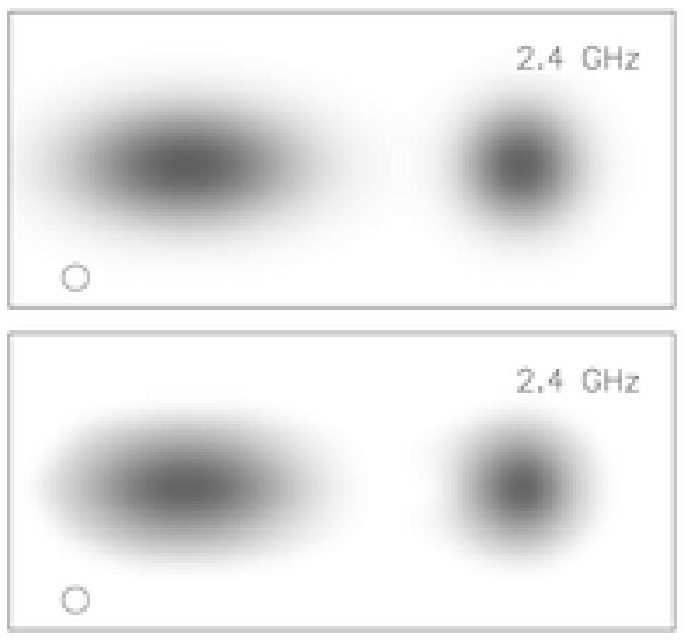,scale=0.4}
\end{center}
\caption[Band-limited Signals : Multi-frequency images]
{\small Band-limited Signals - Multi-frequency images : 
These images show a comparison between the true sky brightness (top row)
and the brightness reconstructed using the MS-MFS algorithm (bottom row) 
at a set of three frequencies (1.0, 1.8, and 2.6 GHz from left to right)).
}
\label{Fig:loop_images}
\end{figure}

\begin{figure}[t!]
\begin{center}
\epsfig{figure=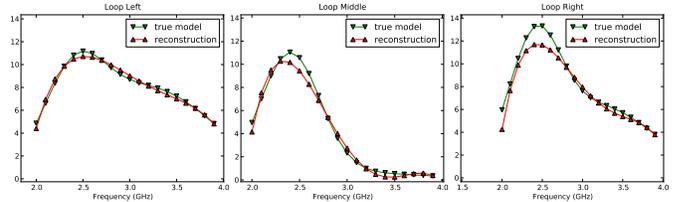,scale=0.27}
\end{center}
\caption[Band-Limited Signals : Spectra across the source]
{\small Band-Limited Signals - Spectra across the source : 
These plots show the true (down-arrows) and reconstructed (up-arrows)
spectra at different locations for the example discussed in this section 
(shown in Fig.\ref{Fig:loop_images}).  The left column corresponds
to the left end of the loop at the location of the leg and shows smooth structure
stretching almost all across the band. The middle column corresponds
to the middle of the source where the only structure in the line-of-sight is
the upper part of the loop (emission at a small fraction of the band).
The right column shows spectra for the brightest point on the
right end of the loop. 
}
\label{Fig:loop_spectra}
\end{figure}

Fig.~\ref{Fig:loop_images} shows
a comparison of the true and reconstructed structure at three different
frequencies (1.2 GHz (left), 1.8 GHz (middle) and 2.6 GHz (right) ).
The top row shows the true structure, and the 
bottom row is the reconstruction. 
The 3D structure is mostly recovered,
with the largest errors being in the central region where the signal spans the shortest
bandwidth. The point source on the right edge of the loop was reconstructed at an angular
resolution slightly lower than that of the highest sampled frequency
and corresponds to the highest frequency at which this spot is brighter than
the background emission.
Fig.\ref{Fig:loop_spectra} compares the true and reconstructed spectra for three
positions in the source (left leg (left), arch of the loop (middle), and right leg (right)).
These spectra show the accuracy to which a $4^{th}$-order
polynomial with MS-MFS was able to reconstruct the structure. 

Spectral-lines are an extreme case of a band-limited signal, and the use of MS-MFS imaging
is restricted to obtaining a wide-band model of the continuum flux from
line-free channels, to be
subtracted out of the data before spectral-line strengths are studied.

\section{Discussion}\label{Sec:CONCLUSIONS}

The introduction of broad-band receivers into radio interferometry has
opened up new opportunities for the study of wide-band continuum
emission from a vast range of astrophysical objects.  
With imaging algorithms that account for the frequency dependence
of the incoming radiation as well as of the instrument, we can 
minimize imaging artifacts, achieve continuum sensitivity and 
reconstruct spatial and spectral structure at the angular-resolution allowed
by the highest observed frequency.

The MS-MFS algorithm models the wide-band sky brightness distribution as a collection
of multi-scale flux components whose amplitudes follow a Taylor-polynomial in frequency.
The data products are a set of
Taylor-coefficient images, which can either be interpreted in terms of continuum intensity, 
spectral index and curvature, 
or used to evaluate a spectral cube, or serve as
a wide-band model for self-calibration or continuum-subtraction. 
For wide-field imaging, multiple pointing-centers (mosaicing)
and the $w$-term are accounted for during
image-formation, and the effect of the primary beam and its frequency-dependence is 
approximately corrected in a post-deconvolution step. 

For point sources for which standard imaging has a dynamic-range limit of a few thousand, this
algorithm achieves dynamic ranges $>10^5$ on test observations with the EVLA, and 
$>10^6$ on noise-free simulations. 
For sources with smooth continuum spectra, it is able to reconstruct spectral information
at the angular resolution allowed by the combined multi-frequency $uv$-coverage.
However, if the visibility functions of very large-scale emission lie mostly within the central
unsampled region of the $uv$-plane, the algorithm requires {\it a-priori} knowledge
about the spectral structure at those scales. 
For high SNR data, error-bars on the spectral-index images range from $<0.02$ 
when multi-scale basis functions are chosen appropriately, 
up to $>0.2$ in the extreme case of modeling smooth extended emission
with a set of $\delta$-functions. In practice,
on-source errors of $\triangle\alpha \approx 0.1$ have been achieved.  
For low SNR extended structure ($<10$), errors due to overfitting can dominate when
a high-order polynomial is used.

There are several directions in which wide-band imaging techniques need to be 
extended and improved. 
The imaging errors in MS-MFS are currently dominated by the multi-scale aspect
of the algorithm, and methods that do adaptive fits \citep{Asp_CLEAN} 
and use more {\it a-priori} information
about large-scale spectra may be more appropriate.
Methods that adaptively find the optimal number of parameters to operate with 
would help in error-control.
Implementations of algorithms such as MS-CLEAN and MS-MFS are inefficient in 
memory-use, and other approaches may be required for large image sizes.
For wide-field imaging, the time variation of the antenna primary beams
must be taken into account during imaging, and wide-band methods combined
with algorithms for direction-dependent corrections 
(A-Projection \citep{AWProjection} or peeling \citep{Smirnov_1_2011}).
For full-Stokes wide-band imaging, where a 
Taylor-polynomial in frequency is not the most appropriate
basis function to model Stokes Q,U,V emission, wide-band imaging with other flux models
must be tried. 

%
%
%


\begin{acknowledgements}
The authors would like to thank the National Radio Astronomy Observatory 
the New Mexico Institute of Mining and Technology, and the Australia Telescope National
Facility for support during
the PhD thesis project that resulted in this algorithm and its implementation
within CASA. The authors would like to thank J.A.Eilek in particular, for extremely helpful 
comments on the presentation of the thesis material that formed the basis
of this paper. 
The authors would also like to thank S.Bhatnagar, K.Golap, R.Nityananda, F.N.Owen, 
R.J.Sault, and M.A.Voronkov, among others, for useful discussions pertaining to this work
and its software implementation.
This project used data from the (E)VLA telescope (test observations and project AR664) operated by
the National Radio Astronomy Observatory, a facility of the National Science Foundation 
operated under cooperative agreement by Associated Universities, Inc.
\end{acknowledgements}


\begin{appendix}
\section{Matrix Notation and Framework}\label{App:A}

The matrix notation used in this paper is explained here, in the context
of an iterative $\chi^2$ minimization process used to solve a system of
linear equations. 
The basic idea is as follows. Let $\A\X=\B$ be the system of equations to be solved
(measurement equations).
The goal is to find a set of parameters $\X$ that minimizes 
$\chi^2 = (\A\X-\B)^{\dag}\W(\A\X-\B)$.
Setting $grad({\chi^2})=0$ to minimize $\chi^2$ leads to a new system of equations called
the normal equations $[\Ad \W \A]  \X = [\Ad]\W\B$. 
The matrix on the LHS is called the
Hessian $[\He]=[\Ad\W\A]$. 
Iterations begin with an initial guess for the parameters $\X$.
These parameters are updated in iteration $i$ as 
$\X_{i+1} \leftarrow \X_i + g [\He^{-1}]\Ad\W (\B-\A\X_i)$, where $g$ controls the step-size.
Iterations continue until a convergence criterion is satisfied.
%
%
The basic iterative framework used
in most imaging and deconvolution algorithms in radio interferometry
can be described using this matrix notation \citep{RAU_IEEE_2009, URV_THESIS}.

\subsection{Measurement Equations}\label{App:measurement}
The measurement equation of an imaging instrument describes its transfer function (the effect of
the measurement process on the input signal).
For an ideal interferometer (a perfect spatial-frequency filter, with no instrumental gains),
it can be written as follows in matrix notation as follows. 
Let $\I^{\rm sky}_{m\times 1}$ be a pixelated image of the sky and let $\V^{\rm obs}_{n\times 1}$
be a vector of $n$ visibilities. 
Let $\Sa_{n\times m}$ be a projection operator that describes
the instrument's sampling function (uv-coverage) as a mapping of $m$ discrete spatial 
frequencies (pixels on a grid) to $n$ visibility samples (usually $n>m$).
Let $\F_{m\times m}$ be the Fourier transform operator.
Then,  the measurement equations become 
$\V^{\rm obs}_{n\times 1} = [{\Sa}_{n\times m}] [\F_{m\times m}] \I^{\rm sky}_{m\times 1} $.

\subsection{Normal Equations} \label{FN:Neqn}\label{App:normal}
The normal equations are the linear system of equations whose solution 
gives a weighted least-squares estimate of a set of parameters in a model  
($\chi^2$ minimization). 
For an ideal interferometer, it is given by 
$[\Fd \Sd \W \Sa \F ] \I^{\rm sky}_{m\times 1} = [\Fd \Sd \W] \V^{\rm obs}_{n\times 1}$ 
where $\W_{n\times n}$ is a diagonal matrix of weights
and $\Sd$ denotes the mapping of measured visibilities onto a spatial-frequency grid.
We can write the Normal equations as $[\He_{m\times m}] \I^{\rm sky}_{m\times 1}= \I^{\rm dirty}_{m\times 1}$
where the Hessian (matrix on the LHS) is by construction a convolution 
operator\footnote{
The convolution of two vectors $\vec{a} \star \vec{b}$ is equivalent to
the multiplication of their Fourier transforms.
A 1-D convolution operator is constructed from $\vec{a}$ and applied to $\vec{b}$
as follows.
Let $[\tens{A}]= diag(\vec{a})$.
Then, $\vec{a} \star \vec{b} =  [\Fd diag([\F]\vec{a}) \F]\vec{b} = [\tens{C}]\vec{b}$.
Here, $[\F]$ is the Discrete Fourier Transform (DFT) operator. 
$[\tens{C}]$ is a Toeplitz matrix, with each row containing a shifted version of $\vec{a}$.
Multiplication of $[\tens{C}]$ with $\vec{b}$ implements the shift-multiply-add sequence
required for the process of convolution. 
}
with 
a shifted version of the point-spread function 
$\I^{\rm psf}_{m\times 1} = diag[\Fd \Sd \W \Sa]$ in each row. 
The dirty image on the RHS is produced by direct Fourier inversion of weighted visibilities.
The normal equations therefore state that the dirty image $\I^{\rm dirty}_{m\times 1}$
is the result of the convolution of $\I^{\rm sky}_{m\times 1}$ with $\I^{\rm psf}_{m\times 1}$.
The solution of these normal equations represents a deconvolution.

\subsection{Iterative solution}\label{App:iterations}
Most existing iterative image reconstruction algorithms in radio interferometry consist
of major and minor cycles.  Major cycles compute the RHS of the normal equations, and
minor cycles perform approximate (implicit) Hessian inversions to calculate updates
for the sky model parameters $\I^{\rm m}$.
The first major cycle starts by transforming the observed visibilities into an image
as $\I^{\rm dirty} = [\Fd\Sd\W] \V^{\rm obs}$, and initializing the sky-model $\I^m$.
Minor cycle steps do a deconvolution to calculate updates to the model $\I^m$.
After several iterations, this updated model is fed into the next major cycle.
This and all subsequent major cycles calculate model visibilities from the 
current model $\V^m = [\Sa\F] \I^m$, calculate
residual visibilities $\V^{\rm res}=\V^{\rm obs} - \V^{\rm m}$, and transform these residuals into 
images $\I^{\rm res} = [\Fd\Sd\W] \V^{\rm res}$.   These residual images replace the initial
dirty image, and a new set of minor-cycle iterations are done.
This process continues until a convergence criterion is reached. Usually, convergence is
defined as $\I^{\rm res}$ being noise-like with no signal left to be extracted in the minor-cycle. 

\section{MS-MFS as implemented in CASA} \label{App:algolisting}

The MS-MFS algorithm described in section \ref{Sec:MSMFS} has been implemented and 
released {\it via} the CASA
software package (version 3.1 onwards).
 More recently, it has been
implemented in the ASKAPsoft\footnote{ASKAPsoft is the software package being
developed at the CSIRO for the ASKAP telescope.} package.
Algorithm \ref{ALGO:MSMFS_1} contains a pseudo-code listing.\\

\noindent The main parameters that control the algorithm are 
\begin{enumerate}
\item $\nu_0$ : a reference frequency chosen near the middle of the sampled
frequency range, about which the Taylor expansion is performed, 
\item $\Nt $ : the number of coefficients of the Taylor polynomial to solve for, and 
\item $\Ns $ and $\I^{\rm shp}_s$ : a set of scale sizes in units of image pixels to use
for the multi-scale representation of the image. In order to always allow for the
modeling of unresolved sources, the first scale function $\I^{\rm shp}_0$ is 
forced to be a $\delta$-function.
\end{enumerate}
The data products are $\Nt $ Taylor-coefficient images, a spectral index image, and a 
curvature image (if $\Nt >2$).
This wide-band image model can be used within a standard self-calibration loop.

\begin{algorithm}
\label{ALGO:MSMFS_1}
  \SetLine
  \linesnumbered
  \dontprintsemicolon
  \KwData{calibrated visibilities : $\vec{V}^{\rm obs}_{\nu}~~\forall \nu$}
  \KwData{$uv$-sampling function and weights : $[\Sa_{\nu}], [\Wimn]$}
  \KwData{input : number of Taylor-terms $\Nt $, number of scales $\Ns $}
  \KwData{input : image noise threshold, $\sigma_{\rm thr}$, loop gain $g$}
  \KwData{input : scale basis functions : $\I^{\rm shp}_s ~ \forall s\in\{0,\Ns -1\}$}
  \KwData{input : reference frequency $\nu_0$ to compute $w_{\nu}=\dnuno$}
  \KwResult{model coefficient images : $\I^{\rm m}_{t}~ \forall t\in\{0,\Nt -1\}$}
  \KwResult{intensity, spectral index and curvature : $\I^{\rm m}_{\nuup_0},\I^{\rm m}_{\alphaup},\I^{\rm m}_{\betaup}$}

  \vspace{0.5cm} 
  \For{ $t \in \{0,\Nt -1\}, q \in \{t,\Nt -1\}$}
  {
        { Compute the spectral Hessian kernel } $\vec{I}^{\rm psf}_{tq} = \sum_{\nu} \wntq \vec{I}^{\rm psf}_{\nu}$\;
        \For{ $s \in \{0,\Ns -1\}, p \in \{s,\Ns -1\}$}
	{
		{Compute scale-spectral kernels} $\vec{I}^{\rm psf}_{{sp}\atop{tq}} = \vec{I}^{\rm shp}_s \star \vec{I}^{\rm shp}_p \star \vec{I}^{\rm psf}_{tq} $\;
	}
  }
  \For { $s \in \{0,\Ns -1\}$}
  {
     Construct $[{\He^{\rm peak}_s}]$ from $mid(\I^{\rm psf}_{{s,s}\atop{t,q}})$ and compute $[{\He^{\rm peak}_s}^{-1}]$\;
  }
  Initialize the model $\vec{I}^{\rm m}_t$ for all $t \in \{0,\Nt -1\}$\; 
  \vspace{0.5cm} 
  \Repeat (\tcc*[f]{Major Cycle}) { Peak residual in $\vec{I}^{\rm res}_0 < \sigma_{\rm thr}$ }
  {
    \For{$t \in \{0,\Nt $-$1\}$}
    {
      Compute $\vec{I}^{\rm res}_t = \sum_{\nu} \wnt \vec{I}^{\rm res}_{\nu}$ from residual visibilities $\V^{\rm res}_{\nu}$\;
      \For{$s \in \{0,\Ns $-$1\}$}
      {
	    Convolve with $s^{th}$ scale-function $\vec{I}^{\rm res}_{{s}\atop{t}} = \vec{I}^{\rm shp}_s \star \vec{I}^{\rm res}_t$
      }
    }
    Calculate minor-cycle threshold $f_{\rm limit}$ from $\vec{I}^{\rm res}_{{0}\atop{0}}$\;
    \Repeat (\tcc*[f]{Minor Cycle}){ Peak residual in $\vec{I}^{\rm res}_{{0}\atop{0}} < f_{\rm limit} $ } 
    {
     \For{$s \in \{0,\Ns $-$1\}$}
     {
       \ForEach{pixel}
       {
          Construct $\I^{\rm rhs}_s$, an $\Nt \times 1$ vector from $\I^{\rm res}_{{s}\atop{t}} ~~\forall ~ t \in \{0,\Nt $-$1\}$\;
          Compute principal solution $\I^{\rm sol}_s = [{\He^{\rm peak}_s}^{-1}] \I^{\rm rhs}_s$\;
          Fill solution $\I^{\rm sol}_s$ into model images $\forall t$ : $\I^{\rm m,sol}_{{s}\atop{t}}$
       }
       Choose $\I^{\rm m}_{{p}\atop{t}} = max\{\I^{\rm m,sol}_{{s}\atop{t=0}},~\forall~s\in\{0,\Ns $-$1\}\}$ \;
     }
       \For{$t \in \{0,\Nt -1\}$}
       {
        Update the model image : $\I^{\rm m}_t = I^{\rm m}_t + g ~[ \I^{\rm shp}_{p} \star \I^{\rm m}_{{p}\atop{t}}]$ \;
        \For{$s \in \{0,\Ns $-$1\}$}
	{
          Update the residual image : $\I^{\rm res}_{{s}\atop{t}} = \I^{\rm res}_{{s}\atop{t}} - g ~\sum_{q=0}^{\Nt -1}[\I^{\rm psf}_{{sp}\atop{tq}} \star \I^{\rm m}_{{p}\atop{t}}]$\;
	}
       }
    }
   Compute model visibilities $\V^{\rm m}_{\nu}$ from  $\I^{\rm m}_t~\forall t\in \{0.\Nt -1\}$\;
   Compute residual visibilities $\V^{\rm res}_{\nu} = \V^{\rm obs}_{\nu}-\V^{\rm m}_{\nu}$\;
  }
  \vspace{0.5cm} 
Restore the model Taylor-coefficients $\I^{\rm m}_t~\forall t\in \{0.\Nt -1\}$ \;
Calculate $\vec{I}^{\rm m}_{\nuup_0}, \vec{I}^{\rm m}_{\alphaup}, \vec{I}^{\rm m}_{\betaup}$ from $\I^{\rm m}_t~\forall t\in \{0.\Nt -1\}$\;
If required, remove average primary beam : $\vec{I}^{\rm new}_{\nuup_0}=\vec{I}^{\rm m}_{\nuup_0}/\vec{P}_{\nuup_0};  \vec{I}^{\rm new}_{\alphaup}=\vec{I}^{\rm m}_{\alphaup}-\vec{P}_{\alphaup};  \vec{I}^{\rm new}_{\betaup}=\vec{I}^{\rm m}_{\betaup}-\vec{P}_{\betaup}$\;

  \caption[MS-MFS Algorithm]
         {MS-MFS, as implemented in CASA}
\end{algorithm}

\end{appendix}

\end{document}